# Uniform microwave field formation for control of ensembles of negatively charged nitrogen vacancy in diamond

Oleg Rezinkin,[1,2] Marina Rezinkina,[1,2,a] Takuya Kitamura,[1] Rajan Paul[1] and Fedor Jelezko[1]

[1] Ulm University, Ulm, 89081, Germany

[2] National Technical University "Kharkiv Polytechnic Institute", Kharkiv, 61002, Ukraine

[a] The author to whom correspondence may be addressed: maryna.rezynkina@uni-ulm.de

## ABSTRACT

The homogeneity of the microwave magnetic field is essential in controlling a large volume of ensemble spins, for example, in the case of sensitive magnetometry with nitrogen-vacancy (NV) centers in diamond. This is particularly important for pulsed measurement, where the fidelity of control pulses plays a crucial role in its sensitivity. So far, several magnetic field-forming systems have been proposed, but no detailed comparison has been made. Here, we numerically study the homogeneity of five different systems, including a planar antenna, a dielectric resonator, a cylindrical inductor, a barrel-shaped coil, and a nested barrel-shaped coil. The results of the simulation allowed us to optimize the design parameters of the barrel-shaped field-forming system, which led to significantly improved magnetic field uniformity. To measure this effect, we experimentally compared the homogeneity of a field-forming system having a barrel shape with that of a planar field-forming system by measuring Rabi oscillations of an ensemble of NV centers with them. Significant improvements in inhomogeneity were confirmed in the barrel-shaped coil.

## I. INTRODUCTION

Negatively-charged nitrogen-vacancy (NV) centers in diamonds have been studied as quantum sensors for various physical quantities, including the magnetic field,[1] the electric field,[2] temperature,[3] and more. In particular, sensitive magnetometry with picotesla sensitivity has been demonstrated using an ensemble of NV centers 500 μm thick, initialized by a green laser focused down to 47 μm.[4] However, increasing the number of controlled NV centers is a difficult problem, since it requires that the controlling magnetic field have high spatial uniformity in volume about fractions of a cubic millimeter. Microwave (MW) electromagnetic fields (EMF) are crucial components in controlling the spin states of the NV centers. Specifically, for Ramsey-type pulsed measurements, their inhomogeneity significantly affects the fidelity of the control pulses and, thereby, the sensitivities of these measurements.[5,6]

So far, several field-forming systems have been reported for this application. Striplines[7,8] or omega-like shapes[9-13] have been designed for small volumes of NV centers. For larger volumes of ensembles, various planar systems have been developed, including planar double loop gap resonators[14-16] and so on.[17-24] However, they are insufficient for a larger ensemble of NV centers because the magnetic field becomes significantly weaker outside the field-forming plane of the system. Therefore, several approaches have been made to create homogeneous fields over a large, 3-dimensional volume, such as the use of dielectric resonators,[25-28] sandwiching two planar systems,[29-31] use of coils,[32,33,] and the like.[34] The dielectric resonators greatly improve the homogeneity of the MW magnetic field. Although its resonant frequency could be tunable, it is questionable if a relatively high Q resonator is compatible with multi-frequency control for some applications, including vector magnetometry[35,36], concatenated continuous dynamical decoupling[37,38], chirped pulses[39], and so on. Furthermore, its integration into a compact device for industrial applications[40,41] is uncertain due to its size and cost. On the other hand, the use of wires or PCBs is compatible with multi-frequency control and cost-effective, but few studies have comprehensively investigated their homogeneity.

The study aims to investigate the homogeneity of different systems by numerical simulation and experimentally compare the 3D systems with a planar antenna system. For this purpose, we focus in this work on the issues of the EMF distribution and not on the frequency characteristics. We first simulate the magnetic field distributions of different systems to compare their uniformities and describe their system-dependent characteristics. They include (A) a planar antenna and a series of 3D systems, including (B) a dielectric resonator, (C) a cylindrical inductor, (D) Helmholtz coils, (E) a barrel-shaped coil, and (F) a nested barrel-shaped coil. For example, the barrel-shaped coil is fabricated, and its homogeneity is tested by measuring the Rabi oscillations of an ensemble of NV centers in a diamond. This is compared with that of a planar field-forming system.

This paper is organized as follows: Section 2 describes the simulation of the five MW field-forming systems and their comparison. Section 3 details the barrel-shaped coil's manufacturing process. Section 4 provides an NV-based experimental comparison of the barrel-shaped coil with the planar structure. Section 5 concludes the paper.

## II. MATHEMATICAL SIMULATION OF THE GIGAHERTZ FREQUENCY MAGNETIC FIELD-FORMING SYSTEMS

To implement sensing methods based on continuous wave optically detected magnetic resonance or driving of coherent oscillations (Rabi oscillations) using an ensemble of NV centers, it is necessary to generate a magnetic field in the MW frequency range close to $f_0 = 2.87$ GHz and to provide a bandwidth of approximately $\Delta f = 20$ MHz.[25] These constraints are due to the zero-field splitting of NV centers, which sits at around 2.87 GHz, and the multiple resonance curves, which have a width of more than 2 MHz and need to be separated. These conditions provide a $Q$-factor $Q = \frac{f_0}{\Delta f} = 143.5$, which is low enough to prevent ringing during pulsed measurements with time constant $\tau_c = \frac{Q}{\pi f_0} = 16$ ns. It is also necessary to consider the possible shift in the operating frequency range associated with a temperature drift and the influence of a possible magnetic background.

For operating an ensemble of NV centers, generating a magnetic field with a high degree of uniformity and a relatively high magnetic flux density in a diamond is necessary.

Depending on the alignment of the optical system, the location of the area of the diamond in which the color centers are excited may shift slightly. Therefore, the exact location of the boundaries of the ensemble within the field-forming system is unknown in advance. This leads to the fact that the field generation system must form such a field distribution in which small displacements of the ensemble boundaries do not significantly change the inhomogeneity of the MW field.

Such a magnetic field (MF) should be distributed in a considered volume occupied by the NV centers with the inhomogeneity level, $\sigma_{pp}$, of no more than 0.5 % (where $\sigma_{pp} = 2 \cdot 100\ \% \cdot (B_{\max 1} - B_{\min 1})/(B_{\max 1} + B_{\min 1})$, and $B_{\max 1}$, $B_{\min 1}$ are the maximal and the minimal levels of the magnetic flux density).[30] The exact geometrical parameters of the volume of diamond containing involved NV centers depend on the optical system used. As a rule, coinciding or oppositely directed exciting green (532 nm) and red (600 – 850 nm) luminescence radiation can be used. We consider an option when these directions coincide. For the optical system we use, the required region of homogeneous MW excitation has the form of a cylinder with length $H_{MF} = 0.5 - 1.2$ mm and diameter $D_{MF} = 50$ μm. The overall dimensions of the diamond hosting the NV centers must also fit into a cylindrical region with a diameter of 3 mm (e.g. a parallelepiped-shaped diamond with dimensions 1 mm×1 mm×0.5 mm). The main focus of this work is devoted to ensuring a high degree of magnetic field homogeneity, and hence the issues of matching field-forming systems and the efficiency of photon collection from NV centers are not considered here. Since the flux density $B$ depends on the current, which in turn depends on the matching parameters, which we do not account for, the below-mentioned values of $B$ are merely for reference. Since the systems under consideration are linear, the magnetic field strength does not affect the degree of its homogeneity. In the COMSOL simulations, the same current, $I_{\max}$, was set in all field-forming systems and was calculated using a fixed power level $P = 10$ W and an impedance of the MW source $Z_{in} = 50$ Ohm, which gives us $I_{\max} = (2 \cdot P/Z_{in})^{1/2} = 0.6325$ A.

To solve the problem of the uniform microwave field formation, we consider several field-forming systems and compare the homogeneity of their magnetic fields. The EMF parameters are obtained with the help of mathematical modeling using COMSOL Multiphysics 5.6. To simplify the simulations, we used a 2D approach with cylindrical symmetry as all considered systems can be approximately presented as cylindrical, having axial symmetry axis $r_0 = 0$. Since the degree of cylindrical asymmetry for all systems considered in this work is relatively small, such an assumption is possible as a first approximation to solving the problem of choosing configurations and parameters of field-forming systems that ensure a high degree of magnetic field homogeneity. In further studies, a more detailed modeling of the selected systems is planned, considering their absolute configuration. In the considered MW range, the EMF source was modeled as a sinusoid current through the windings of the field-forming system. As the considered inductors' overall dimensions are relatively small (about 3 mm × 3 mm × 3 mm) compared to the considered EMF wavelength $\lambda = 104.53$ mm, the problem of the magnetic field calculation can be solved in the quasi-static approximation. In such a case, windings with the specified currents can be presented as coils in the AC/DC module of COMSOL Multiphysics. To consider wave processes, the phase lags were assigned for currents of each multi-turned inductor winding with the radius $R$ as follows: $\varphi_R = (2\pi R/\lambda)2\pi$ [rad].

The same current level, $I_{\max}$, was used to compare the magnetic flux levels and the degree of the MF inhomogeneity in the inner cylindrical domain (ICD), having length $H_{MF}$ and diameter $D_{MF}$. The axis of symmetry of ICD coincides with the axis of symmetry of field formation systems. To quantify the degree of the MF inhomogeneity, we will use the parameter $\sigma_{pp}$.[30] The results of $\sigma_{pp}$ levels for the considered field formation systems (described below) are presented in Fig. 1 and Table I.

In addition to the distributions of magnetic field inhomogeneity in the systems under consideration, shown in Fig. 1, the results of magnetic field modeling are presented in the form of spatial distributions in a relatively large area surrounding the field-forming systems under study (see Figs. 2(b), 3(b), 4(b), 5(b), 6(b)), as well as in a relatively small area directly adjacent to the axis symmetry of systems and including the zone with the NV centers (see Figs. 2(c), 3(d), 4(c), 5(c), 6(c)). This representation of information about the distribution of magnetic field induction allows us to display exactly how the magnetic field is distributed in the system as a whole. The change in the density of the magnetic field induction and the intensity of this change along the spatial axes in NV containing zone is also shown. Such a representation permits us to take into account how the configuration of a field-forming system changes the magnetic field in the entire system, as well as the magnitude of the induction and its gradient in the local zone with NV centers.

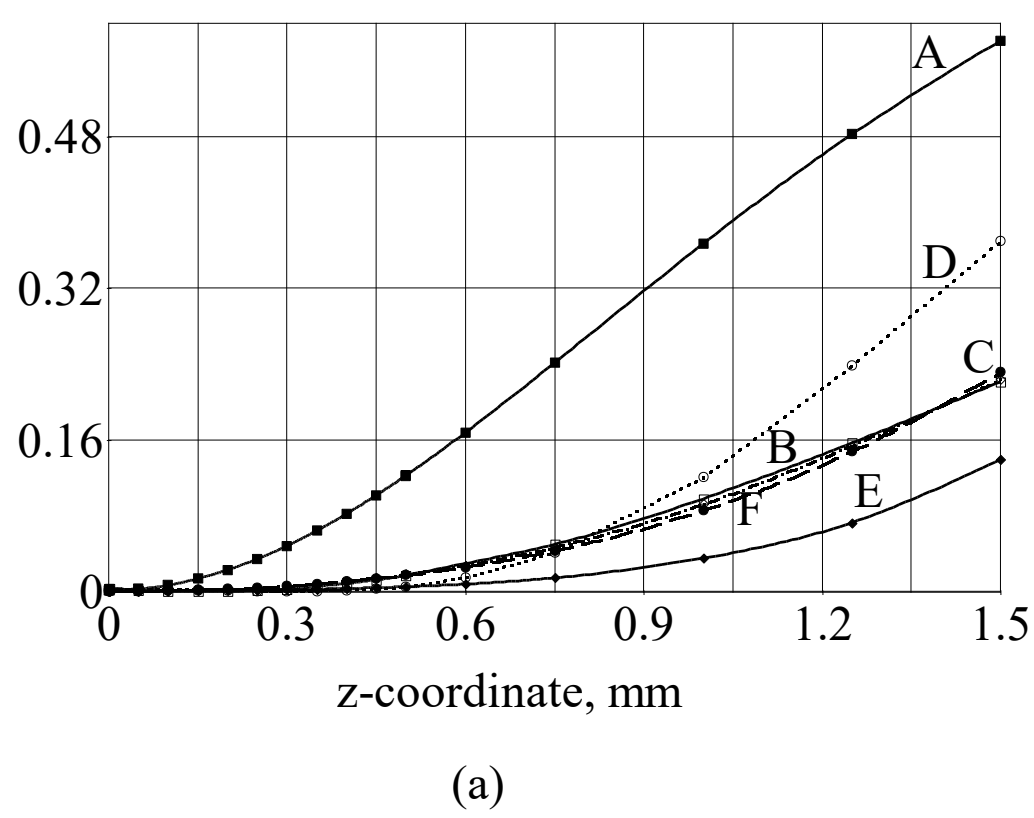

(a)

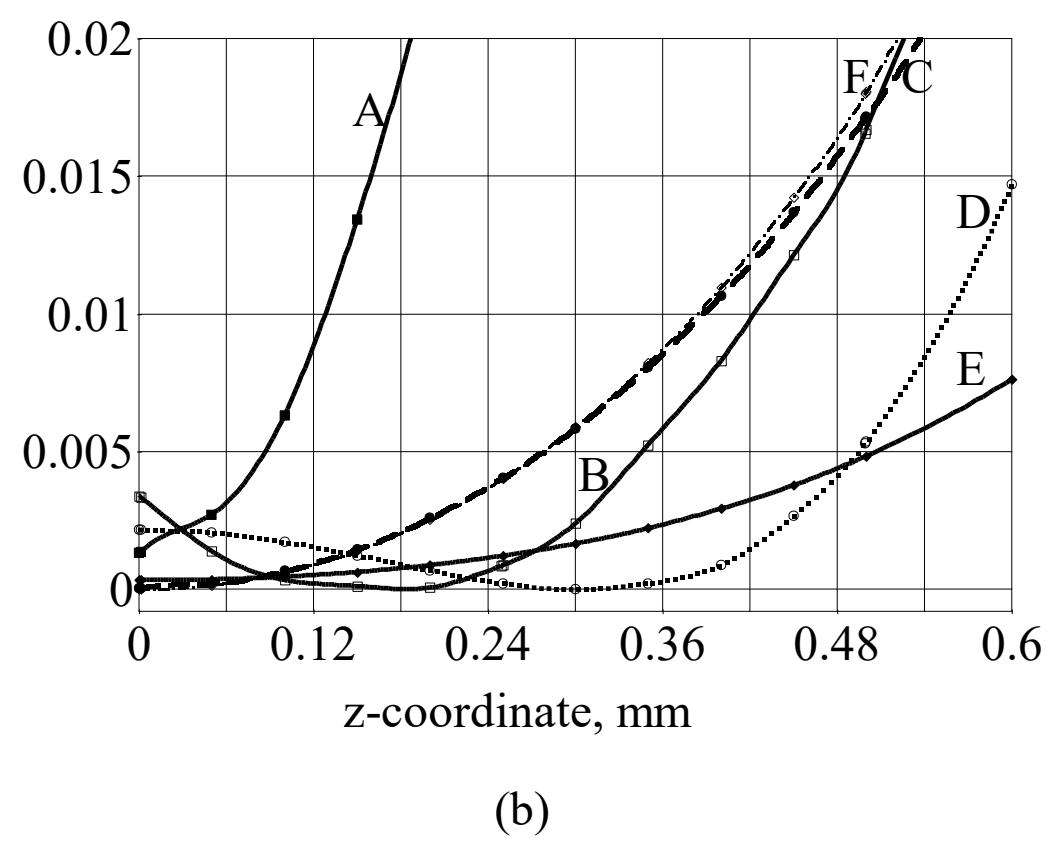

(b)

FIG. 1. Calculated distributions of the inhomogeneity level $\sigma_{pp}$ along the $z$ axis at $r$=0: (a) – in the range 1 – 1.5 mm, (b) – in the range 1 – 0.6 mm. Geometries numbers A, B, C, D, E, and F correspond to the Table I and Figs. 2(a), 3(a), 4(a), 5(a), 6(a), 7(a).

TABLE I. Calculated parameters of the field-forming systems

| Field-forming geometry | $Nw$ | $D$×$H$, mm | $\sigma_{pp}$ in ICD, % $\|z\|$<0.25 mm | $\sigma_{pp}$ in ICD, % $\|z\|$<0.6 mm |
|---|---|---|---|---|
| A* | 1 | 7.8×0.018 | 3.4 | 20 |
| B** | 1 | 30×3 | 0.34 | 2.9 |
| C | 3 | 3×3 | 0.4 | 2.6 |
| D | 2 | 3×1.5 | 0.2 | 1.49 |
| E | 2×3 | 4×3 | 0.1 | 0.735 |
| F | 4×3 | 5×4 | 0.4 | 2.8 |

$z$ – azimuthal coordinate
* - planar MW antenna
** - dielectric resonator with ε=200
$N_w$ is the number of windings of the field-forming system;
$D$×$H$ is the overall diameter and the height of the field-forming system.

## A. Planar MW antenna

Let us consider an approach to solving the problem of homogeneous MW MF generation with the help of a planar MW antenna, which has been well described elsewhere.[11] The peculiarity of this system is that its impedance is close to $Z_{in}$. So, it is self-matching with the source, and the connection of the additional lumped elements for matching is redundant. Flat single-turn field-forming systems, similar to this one, are widely known and are used to excite ensembles of NV centers formed at or very near the surface of the host diamond.[9,16,18] The results of the EMF simulation in a system with such a configuration are shown in Figs.: 1, 2, and Table I (geometry A). Fig. 2(a) shows a schematic of the planar MW antenna made from copper and with the following parameters: outer diameter $D_{out}$ = 8 mm, inner diameter $d_{in}$ = 3 mm, thickness $w_w$ = 18 μm.

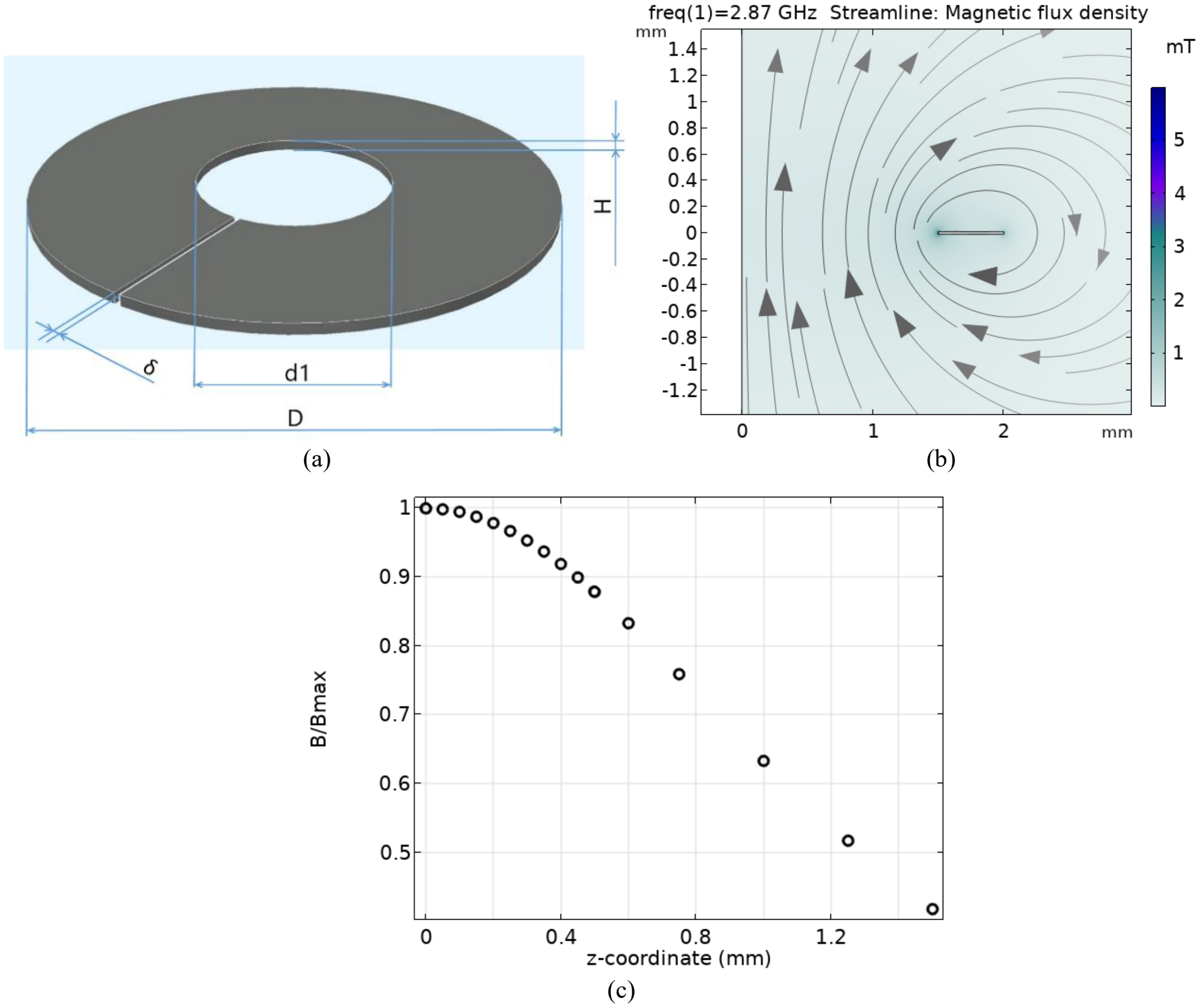

FIG. 2. (a) – sketch of the system (geometry A in Table I) $D = 7.8$ mm, $H = 0.018$ mm, $d_1 = 3$ mm, $\delta = 100$ µm; (b) is the calculated magnetic flux density in the calculation domain; (c) – calculated values of the relative magnitude of the magnetic flux along the $z$ axis at $r$=0 ($B_{max}$ is magnetic flux density at the center of the field-forming system at $z$=0, $r$=0).

As can be seen from Fig. 1 and Table I (geometry A), Fig. 2, MF with a homogeneity level of 0.5 % is located only in a relatively small area $|z| < 0.075$ mm, less than that required for an ensemble of NV centers.

### B. Dielectric resonator

We now consider a series of field-forming systems designed to extend the homogeneous field region into 3-dimensions to allow a larger ensemble of NV centers to be controlled. The known approach to this problem solution is the usage of the dielectric resonator (see, for example[25-28]). Fig. 3(a) shows a schematic of such a resonator with following parameters: diameter of dielectric $D$=5 mm, height $H = D{\cdot}0.5 = 2.5$ mm, inner diameter $d$=3 mm, thickness of the dielectric board $d_{wF} = 200$ µm, coil section diameter $d_w = 150$ µm, relative permittivity of the resonator dielectric $\varepsilon_r = 200$, relative permittivity of the dielectric board: $\varepsilon_{vaf} = 3.55$, (RO4003 Rogers Corporation). The results of the MF simulation in such a system are shown in Fig. 3(b), 3(c), and 3(d). These parameters were chosen due to a series of calculations performed with various resonator geometries and $\varepsilon_r$ values to achieve higher levels of MF.

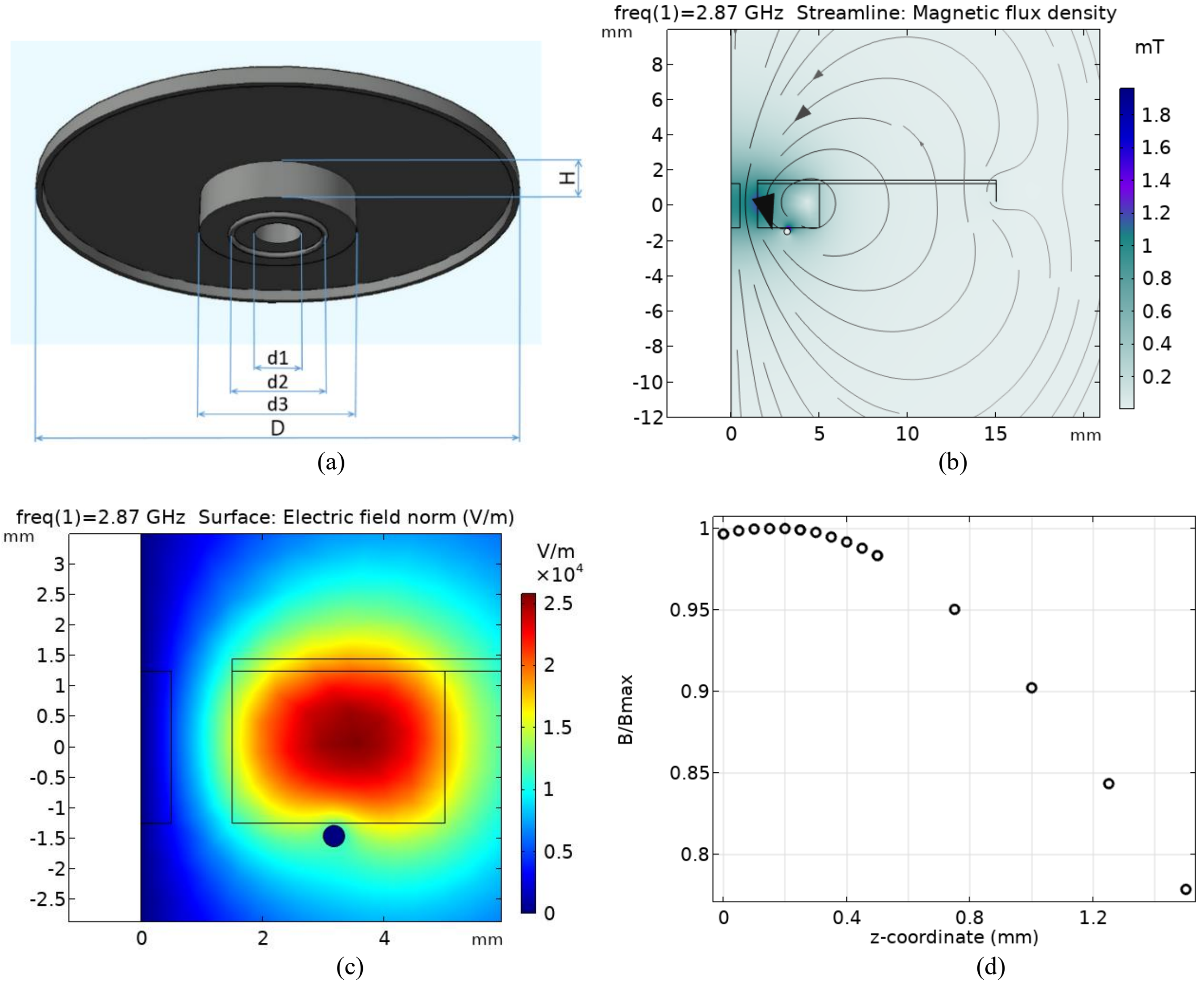

FIG. 3. (a) – sketch of the system (geometry B in Table I) $D$ = 30 mm, $H$ = 3 mm, $d_1$ = 3 mm, $d_2$ = 6 mm, $d_3$ = 10 mm; (b) – calculated magnetic flux density in the calculation domain; (c) – calculated electric field intensity in the calculation domain; (d) – calculated values of the relative magnitude of the magnetic flux along the $z$ axis at $r$=0 ($B_{max}$ is magnetic flux density at the center of the field-forming system at $z$=0, $r$=0).

As can be seen from the simulations (see Fig. 1 and Table I geometry B, Fig. 3), the level of the inhomogeneity is 0.5 % in the area $|z| < 0.35$ mm.

**C. Cylindrical inductor**

Let us consider a cylindrical inductor consisting of three turns of strips made of copper foil (see Fig. 4(a)). Its geometrical parameters are as follows: height $H$ = 3 mm, inner diameter $d_{in}$ = 3 mm, and the thickness of the copper foil $w_w$ = 15 µm. Insulation between turns of the copper foil is made of polyimide film (Kapton) with thickness $w_{in}$ = 65 µm.

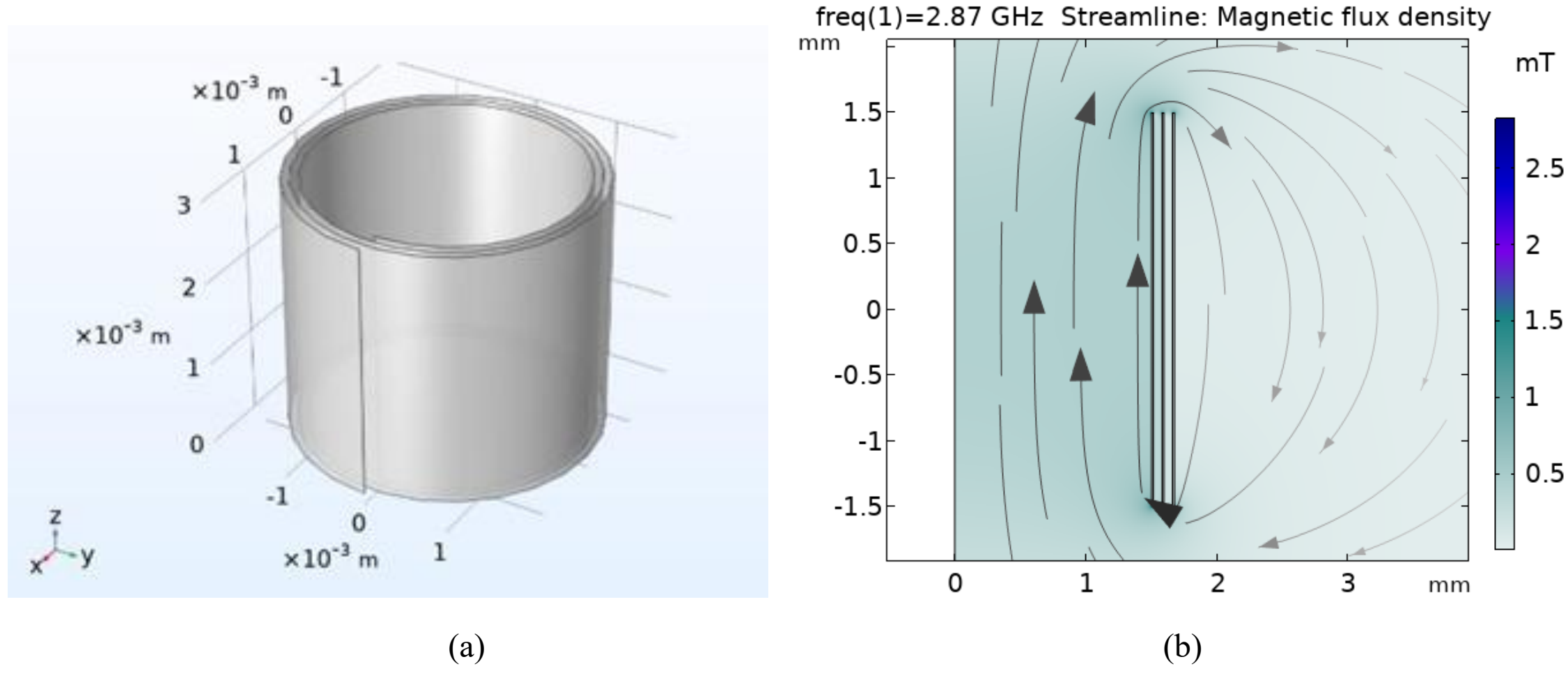

(a) (b)

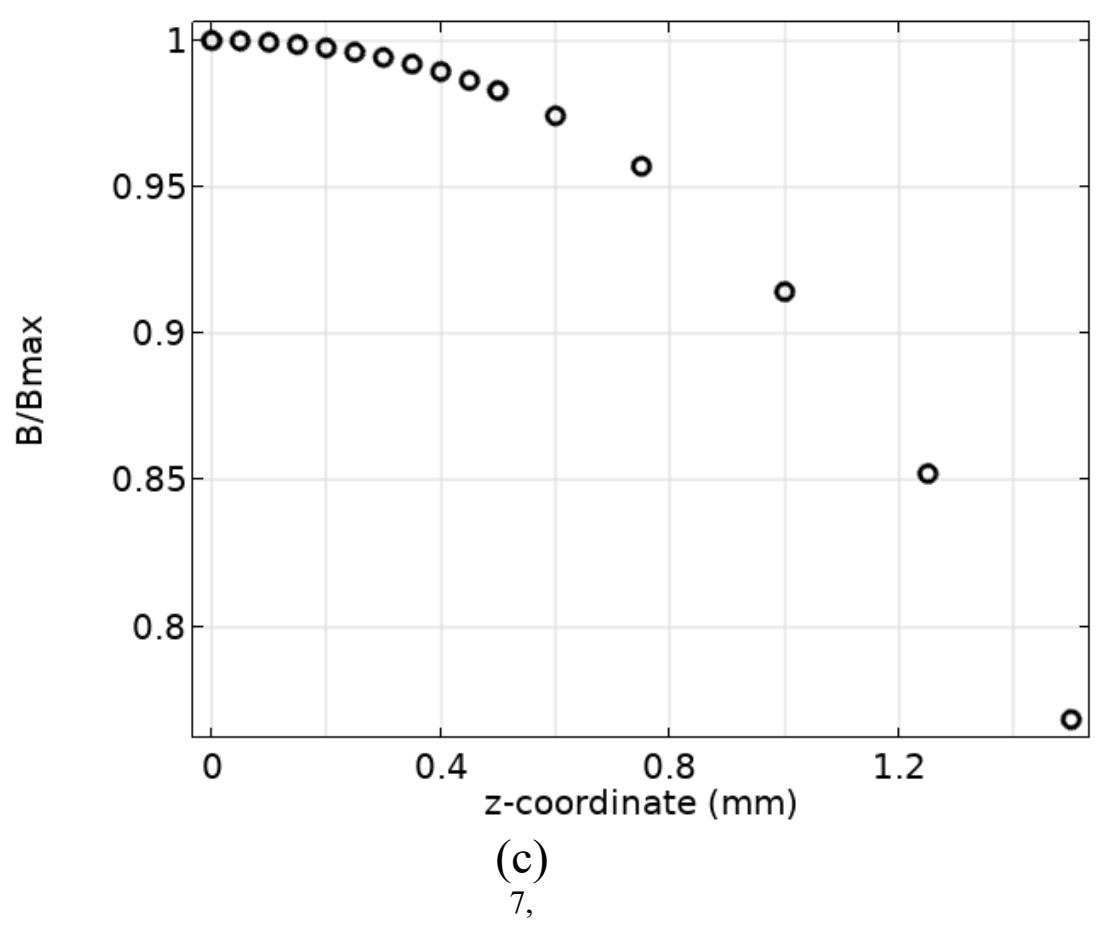

(c)

7,

FIG. 4. (a) – sketch of the system (geometry C in Table I); (b) – calculated magnetic flux density in the calculation domain; (c) – calculated values of the relative magnitude of the magnetic flux along the $z$ axis at $r$=0 ($B_{max}$ is magnetic flux density at the center of the field-forming system at $z$=0, $r$=0).

As simulations show (see Fig. 1 and Table I geometry C, Fig 4), this inductor provides the magnetic field inhomogeneity of 0.5 % in the area $|z| < 0.3$ mm.

**D. Field-forming system in the form of Helmholtz coils**

One of the configurations of field-forming systems that provide a uniform electromagnetic field is Helmholtz coils.[29 – 32] To study the degree of homogeneity of the magnetic field in the ICD, its simulation was carried out in such a system. The geometrical parameters of the Helmholtz coils are as follows: $H$=3 mm, $D_{in}$=1.5 mm, and the diameter of the coils $d_w$ = 252 µm. Fig. 1 and Table I (geometry D), Fig. 5, show the results of the magnetic field simulations for this system.

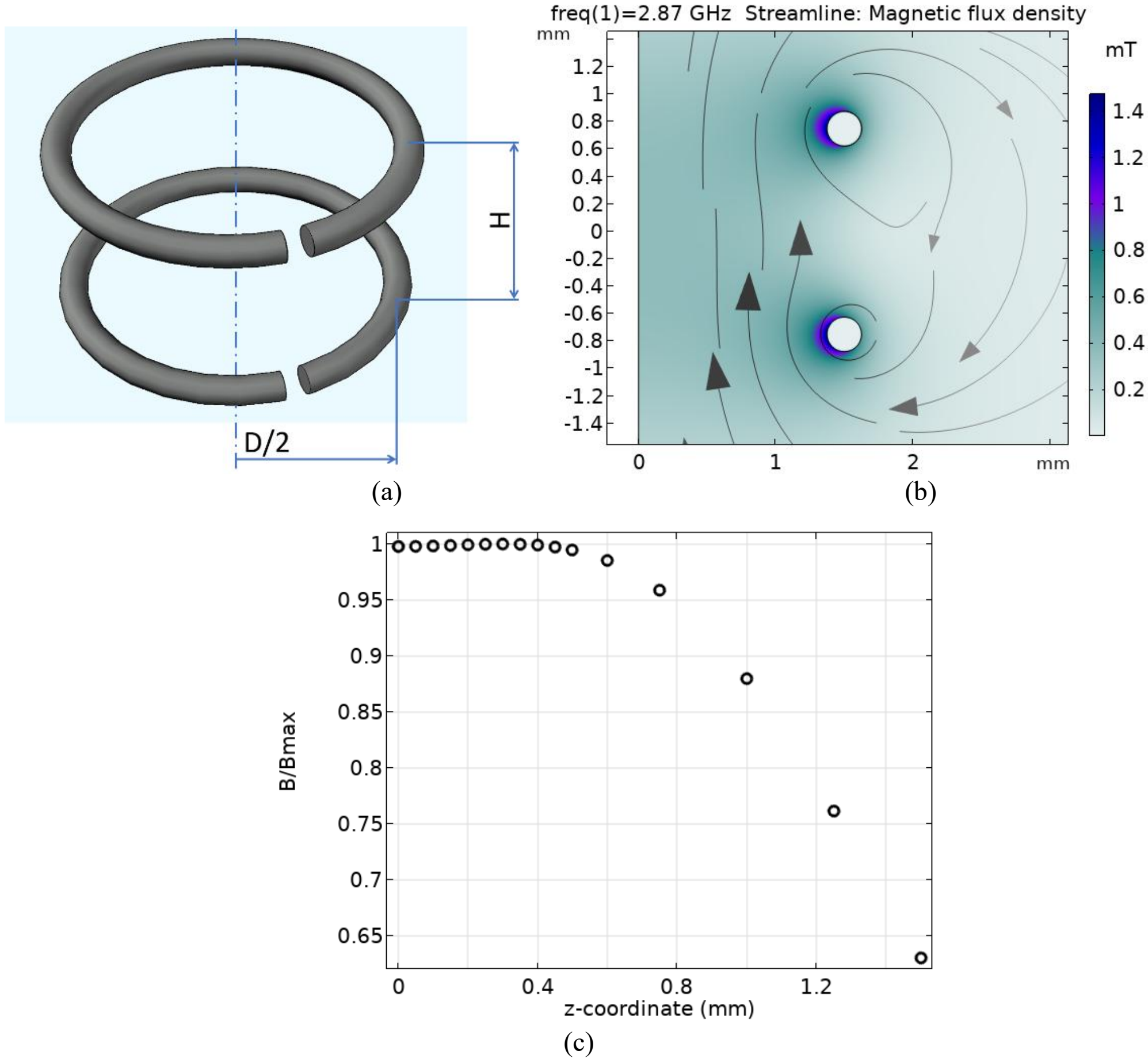

FIG. 5. (a) – sketch of the system (geometry D in Table I) $D$ = 3 mm, $H$ = 1.5 mm, $d_w$ = 252 µm; (b) – calculated magnetic flux density in the calculation domain; (c) – calculated values of the relative magnitude of the magnetic flux along the $z$ axis at $r$=0 ($B_{max}$ is magnetic flux density at the center of the field-forming system at $z$=0, $r$=0).

As simulations show, this inductor provides the magnetic field inhomogeneity of 0.5 % in the area $|z| < 0.5$ mm.

**E. Field-forming system in the form of a barrel-shaped coil**

In this paper, we also investigate multi-turn coils (see Fig. 6(a)) as a means to increase the magnetic flux density[32] and its homogeneity. The field-forming system has a quasi-axisymmetric configuration formed by two conical windings with large bases directed towards each other. In Fig. 6(a), the terminals of the first winding are designated 1 and 3, and the second are designated 3 and 4. The electrical connection of these conical windings is parallel: terminals 1 and 2 and terminals 3 and 4 are connected in pairs and connected to the source poles. Thus, two three-turn conical windings together form a six-turn barrel-shaped field-forming system. As the simulated model shows, the magnetic field of a similarly-sized multi-turn field-forming system with equal diameter turns connected in series has a lower homogeneity of the magnetic field in the area of interest than the proposed barrel-shaped system. In addition, such a system has a very large inductive reactance, which leads to the need for a source with a high output voltage level.

The influence of the relationship between self-induction and mutual induction of parallel-connected parts of a divided winding leads to a complex correlation between the flux linkages of these parts and the equivalent inductance of the field-forming system. The results of numerical modeling and experimental studies show that the effect of such separation is very significant. In addition to reducing the inductance of the field-forming system, creating a current path from parallel-connected wires, in contrast to strip conductors, makes it possible to organize an optimal distribution of current density in the cross-section of the current conductor. The use of multi-turn systems of wires of various diameters for this optimization will further improve our results.

The geometrical parameters of this inductor are as follows: $H$=3 mm, $d_{in}$=3 mm, wire diameter of the windings $d_w$ = 252 μm, and the total number of turns $N_w$ = 3×2=6. Fig. 1, Table I (geometry E), and Fig. 6 show the results of the magnetic field simulations for this system. This inductor provides the magnetic field inhomogeneity of 0.5 % in the area $|z| < 0.5$ mm.

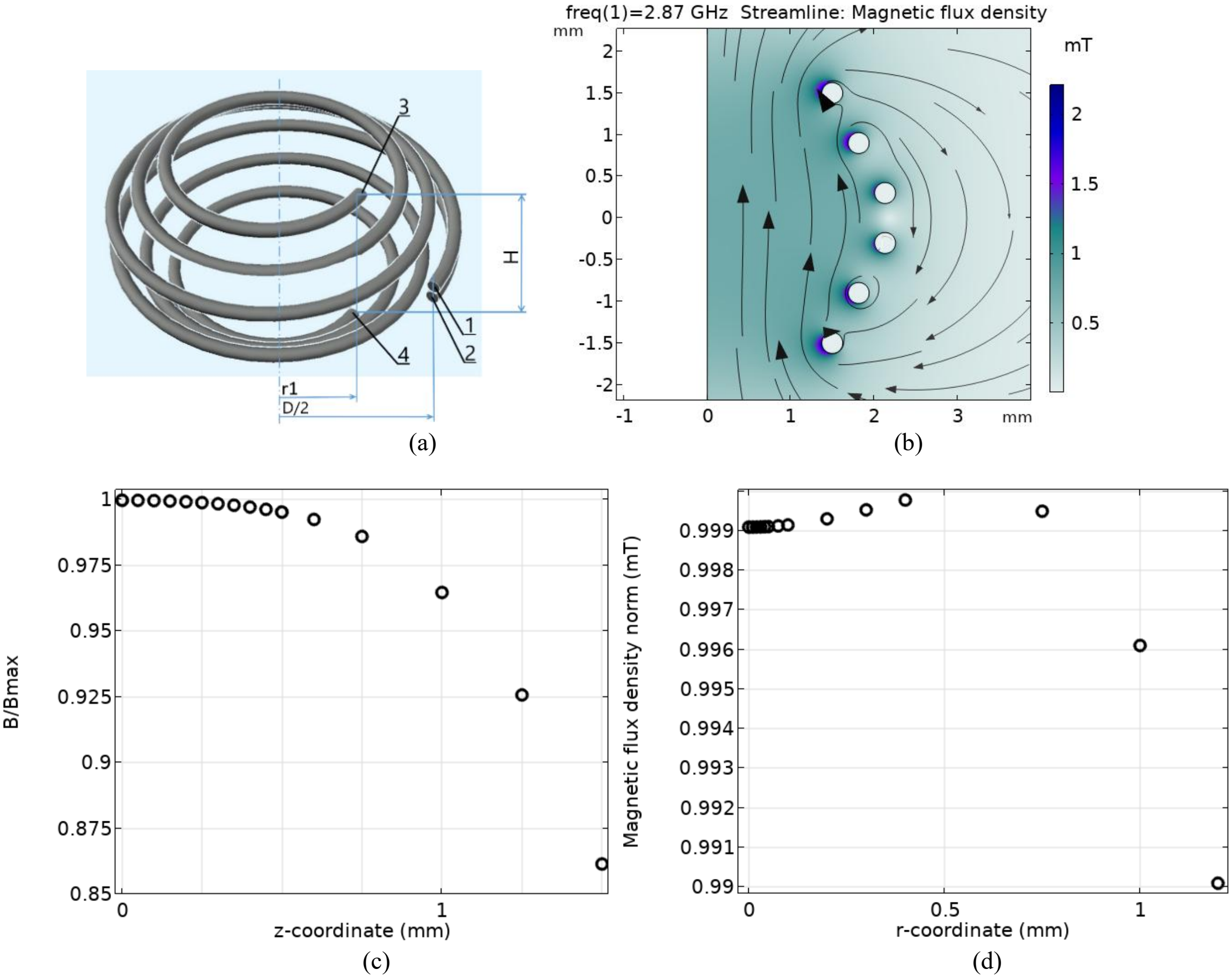

FIG. 6. (a) – sketch of the system (geometry D in Table I) $D$ = 4 mm, $H$ = 3 mm, $r_1$ = 1.5 mm; (b) – calculated magnetic flux density in the calculation domain; (c) – calculated values of the relative magnitude of the magnetic flux along the $z$ axis at $r$=0; (d) – calculated values of the relative magnitude of the magnetic flux along the $r$ axis at $z$=0 ($B_{max}$ is magnetic flux density at the center of the field-forming system at $z$=0, $r$=0).

This inductor provides a high degree of magnetic field uniformity both in the azimuthal (see Fig. 6(c)) and radial (see Fig. 6(d)) directions. Comparisons with other solution[26] obtained by the use of specific techniques to increase the magnetic field uniformity show that the field-forming system proposed in this work provides a significantly more uniform magnetic field. The maximum differences in magnetic flux densities in the field-forming system described in the other work[26] are 1.25% in a small area along the *z*-axis and 10% in a larger area.[26] In the six-coil system proposed in this work (see Fig. 6(a)), such differences are 0.08% and 0.5%, respectively, relative to the scale factor of the geometric dimensions of the field-forming system.. As for the inhomogeneity of the magnetic field in the radial direction, the maximum differences in magnetic flux density in the systems described elsewhere[26] are 3% in a small region along the *r*-axis and 10.5% in a larger region.[26] In our proposed six-turn system, such differences in the first case do not exceed 0.05%, and in the second – 0.1%. Thus, the configuration of the field-forming system proposed in this work allows us to increase the degree of uniformity of the magnetic field significantly.

**F. Inductor with nested barrel-shaped windings**

Magnetic flux density and magnetic field uniformity can be further increased using an advanced version of the previous inductor (see Fig. 6(a)). In this case, the field-forming system consists of four conical windings. Moreover, one smaller pair of conical windings is located inside the other. The electrical connection of all four conical windings is parallel. The geometrical parameters of this inductor are as follows: $H$=3 mm, $d_{in}$=3 mm, wire diameter of the windings $d_w$ = 252 μm, and the total number of turns $N_w$=3×4=12. Fig. 1, Table I (geometry F), and Fig. 7, show the results of the magnetic field simulations for this case. This inductor provides the magnetic field inhomogeneity of 0.5 % in the area $|z| < 0.275$ mm.

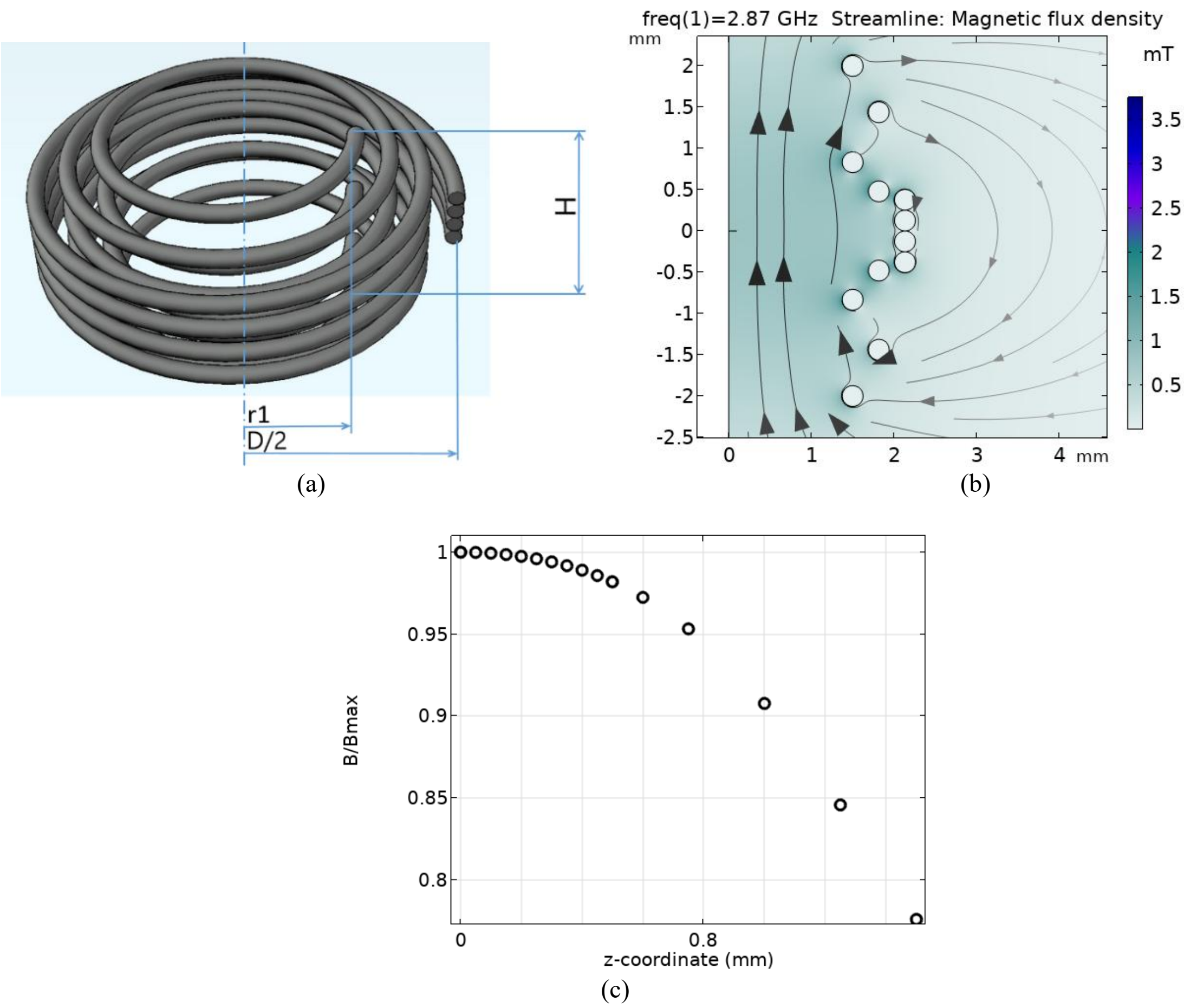

Fig. 7. (a) – sketch of the system (geometry F in Table I) $D$ = 5 mm, $H$ = 4 mm, $r_1$ = 1.5 mm; (b) – calculated magnetic flux density in the calculation domain; (c) – calculated values of the relative magnitude of the magnetic flux along the $z$ axis at $r$=0 ($B_{max}$ is magnetic flux density at the center of the field-forming system at $z$=0, $r$=0).

## III. EXPERIMENTAL MODEL

Prototypes were made to physically model the investigated field-forming systems, and experimental studies were carried out. To realize a physical model of the barrel-shaped coil (geometry E, Table I), a field-forming system with a volumetric quasi-axisymmetric configuration with a working volume of 3 mm in diameter and 3 mm in height was created. A parallelepiped-shaped diamond 1 mm×1 mm×0.5 mm is located on the system's axis in the center of its working area. The physical model of the field-forming system is made of an adhesive polyimide tape (Kapton) 6 mm wide and 60 µm thick (including an adhesive layer) with conductive elements – copper wires with a diameter of 252 µm glued onto it. The conductive part consists of two, three-turn wire spiral elements connected in parallel, forming a six-turn solenoidal inductor system. The photographs show the successive stages of manufacturing the system (see Fig. 8(a)).

On one side (in photo Fig. 8(a) on the right) the wires are soldered together (see terminals 1, 2 in Fig. 6(a), Fig. 8(a), 8(b)), and on the other side (see terminals 3, 4 in Fig. 6(a), Fig. 8(a), 8(b)) they are short-circuited by a piece of copper wire with a diameter of 230 µm. Then, the resulting wire triangle is fixed on the adhesive layer and, together with a polyimide tape, is wound into a roll of three turns. With this winding, the short side of the wire triangle is placed inside, and the acute angle formed by the soldered ends of the wires is placed outside (see Fig. 8(b)). This outer part of the resulting complex winding is one of the current leads and is soldered to the central conductor of an SMA-type coaxial connector (see Fig. 8(b)). The two free ends of the wire short-circuiting the inner ends of the windings are also current leads and are soldered to the outer conductors of the SMA coaxial connector (Fig. 8(b)). When winding in a roll, each subsequent turn has a diameter greater than the previous one by the sum of the polyimide film's thickness and the diameter of the copper wire. Therefore, the spirals of wire generally form a barrel-shaped winding. The two halves of this winding consist of three turns in series, and these halves are connected in parallel to the pins of the SMA connector.

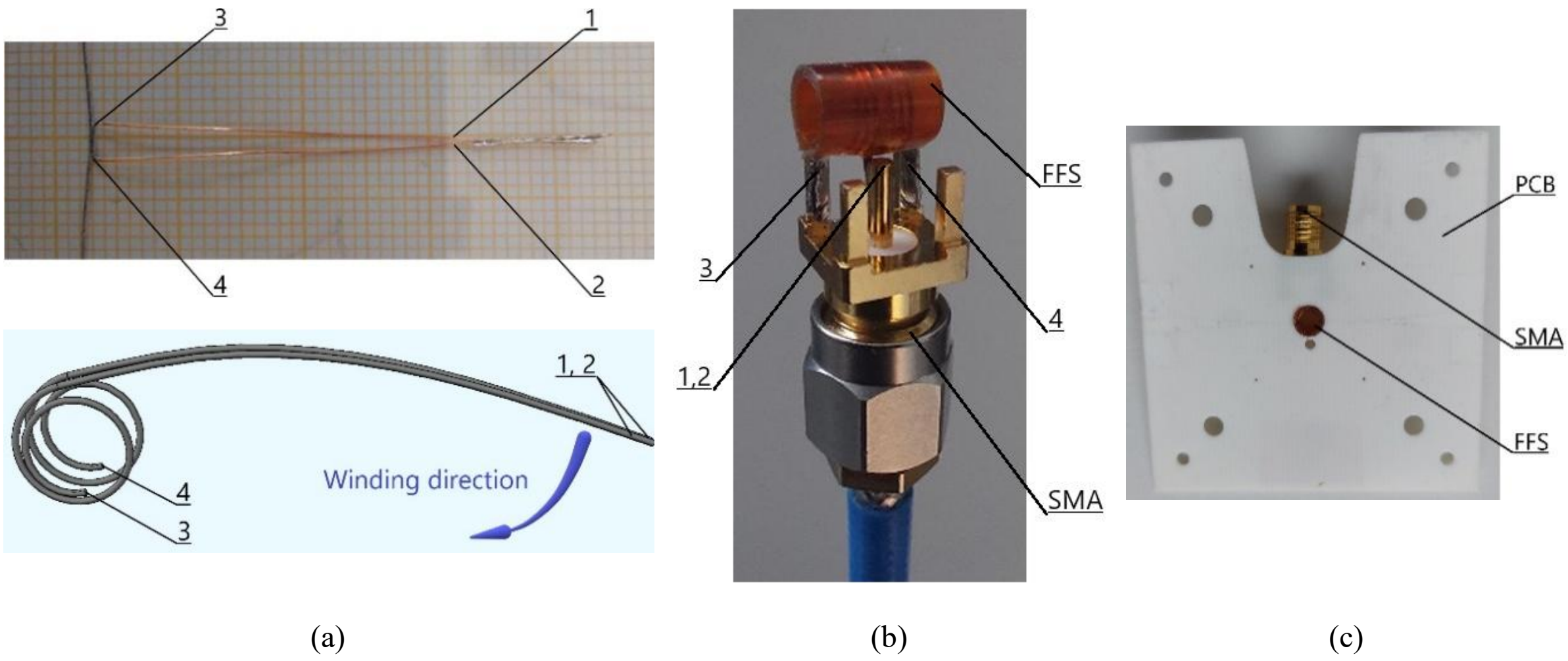

FIG. 8. Model of the MW inductor for excitation of an ensemble of NV centers. FFS – field formation system, SMA – coaxial connector, PCB – printed circuit board, 1–4 terminals (see also Fig. 5(a)). (a) – the system manufacturing, one square in the photo is 1 mm × 1 mm; (b) – connection of the field formation system to the SMA connector; (c) – PCB with the installed field generation system.

The barrel-shaped contours of the resulting complex winding allow for high uniformity of the magnetic field's spatial distribution in the field-forming system's working area. Splitting the current into two identical parallel branches significantly reduces the winding's inductance and resistance.

The MW field distribution of the physical model is investigated by measuring Rabi oscillations using the optical system described in the Appendix. To position the inductor in the optical system along with a coaxial connector, it is soldered to a PCB made of 0.020" (0.508 mm) thick RO4003 (Rogers Corporation) dielectric material with dimensions of 50×50 mm$^2$ (see Fig. 8(c)). The field-forming system is located in the center of the printed circuit board. The design of the experimental setup for testing the manufactured model of the MW structure provides the ability to use micro-screws to displace the printed circuit board relative to the optical axis of the system while leaving the diamond with the ensemble of NV centers motionless. This made it possible to measure the spatial distribution of the magnetic flux density in the working volume of the field-forming system by using the excellent magnetic field sensing capabilities of the NV centers. The design of the PCB makes it possible to place elements to match the input impedance of the inductor with the feeder close to the field-forming wire turns.

Matching the input impedance of the field-forming system and the output impedance of the feeder is important to ensure full utilization of the energy of the MW source. This problem is especially acute when creating mobile devices, in which the mass of MW power sources should be small. One approach to solving the problem of impedance matching is to use resistive broadband matching loads that absorb MW passing through the conductors of the field-forming system, as well as MW antennas that ensure the fulfillment of resonance conditions in the field-forming system itself and do not create reflected waves directed back to the feeder. The second option is to minimize the required power of the MW source. To solve the problem of matching high-frequency devices with a feeder, circuit modeling methods based on the synthesis of an equivalent circuit of a field-forming system consisting of lumped elements are currently widely used. Specific values of the parameters of the equivalent circuit of the field-forming system and elements intended to ensure matching are determined and then verified based on experimental studies of S-parameters using a vector analyzer. However, before solving the problem of matching with the feeder, it is necessary to enable the condition that the distribution of the MW field in the region of the ensemble of NV centers satisfies the specified requirements for its homogeneity. In addition to the problems of configuring the MW field distribution in diamond and impedance matching, there is also a wide range of practically important issues associated with collecting luminescent radiation from NV centers. In this work, we focused on methods of creating a spatial distribution of the MW field in the region of an ensemble of NV centers with uniformity sufficient for use in sensitive magnetic field sensors and other quantum devices that require synchronous spin manipulation. The problems mentioned above are topics for future investigations.

To realize the physical model of geometry E (see Table I), the field-forming system can be produced similarly to geometry E (Table I), but with the number of wire conductors connected in parallel increased from two to four. These wires should also be glued to a polyimide tape, turned into a roll, and connected to a coaxial feeder.

The described technology for manufacturing physical models of geometries D and E (Table I) is convenient for experimental studies of the influence of the field generation system parameters on the distribution of the magnetic field in the working area. In practically important cases, the design of the field generation system can be adapted to the use of known modern technologies, for example, based on the technology of flexible printed circuit boards on a polyimide base with a nickel-gold coating, and for sensitive magnetometers with a copper and silver coating.

Using a vector network analyzer, experimental studies of the scattering parameters of a prototype of the proposed field-forming system (see Fig. 8) were carried out in the range from 2 GHz to 4 GHz. At a frequency of 2.87 GHz, the measured value of $S_{11}$ at the input of the SMA connector to the field-forming system under test is $0.94\angle -115^0$. Simulations carried out using the SimNEC software showed that this field-forming system can be matched using a two-element matching circuit.[42]

## IV. EXPERIMENTAL RESULTS ON THE MAGNETIC FIELD INHOMOGENEITY

To investigate the level of the magnetic field homogeneity of the barrel-shaped coil (geometry D in Table I) – see Fig. 8, Rabi oscillations were measured with the help of the ensemble of NV centers at different positions inside the inductor. For comparison with a planar system, similar experiments have been conducted with a planar lens system.[18] The details of the experiments can be found in the Appendix. Fig. 9(a) shows the Rabi oscillations for the case of the barrel-shaped coil at its center (red circles) and 1.5 mm away in the axial direction from the center (blue squares). In comparison, Fig. 9(b) shows those for the case of the planar system on its surface (red circles) and 1 mm above the surface (blue squares). It is worth noting that during the experiments, the MW powers transmitted to the planar system and to the barrel-shaped coil were different. So the Rabi frequencies do not correspond to the $B_{max}$ levels given in Table 1. The signals in Fig. 9(b) decay relatively quickly, attributed to the magnetic field inhomogeneity. Compared to the planar system, more oscillations are observed, which indicates that the barrel-shaped coil has a more homogeneous field. A comparison of the Rabi periods at different positions shows that the fields from the planar system quickly decrease when away from the board.

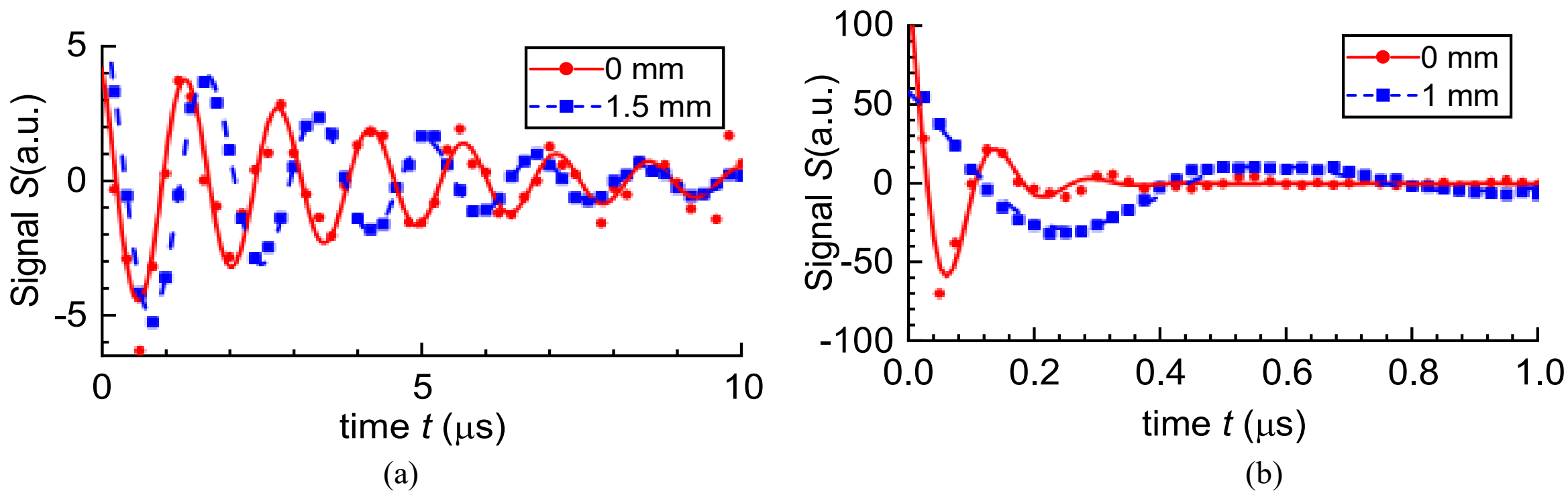

FIG. 9. Rabi oscillations with 500 µm thick NV diamond sample (see Appendix A). (a) – Measurements in the barrel-shaped coil (see geometry D in Table I, Fig. 8). The measurements at the center of the coil and 1.5 mm away in the axial direction are shown as red circles and blue squares with their fit curves shown in the same way. (b) – Measurements in the planar structure at different diamond positions. The measurements on the surface of the PCB board and 1 mm above it are shown as red circles and blue squares, respectively. Likewise, their fit curves with exponentially decaying sinusoidal functions are plotted with a red line and a blue dotted line correspondingly.

We quantify the degree of inhomogeneity by fitting the data with exponentially decaying cosine functions, which can be understood as the integration of Lorentzian-distributed cosine waves.[43] The ensemble averaged signal is modeled as the integration of Rabi oscillations whose frequency is given by $\Omega(\zeta) = \Omega_0(1+\zeta)$, where $\zeta$ is the dimensionless variable characterizing the deviation from the nominal Rabi frequency $\Omega_0$. By assuming that its probability distribution $p(\zeta)$ is given by a Lorentzian function with width $\Delta\zeta$, the integral of the Rabi oscillations can be written as follows:

$$S = Ae^{-t/T_\mathrm{R}} \cos \Omega_0 t,$$

where $T_R = \frac{1}{\Omega_0 \Delta\zeta}$. Therefore, we can obtain the width, $\Delta\zeta$, from the Rabi frequency $\Omega_0$ and the decaying time $T_R$ obtained in the fitting. We associate this width, among other factors, with the inhomogeneity of the magnetic field flux density $B$ by using the correlation (see, for instance[44]):

$$\Omega = \gamma B,$$

where $\gamma$ is the gyromagnetic ratio. It is worth noting that this treatment is based on three assumptions. Firstly, the Rabi oscillations are assumed to be described by cosine waves. This is not true for the spins detuned from the driving frequency. Secondly, the probability distribution of $\zeta$ is assumed to be Lorentzian. This is expected to be system-dependent and is not necessarily Lorentzian. For simplicity, however, we fit the results with exponentially decaying cosine functions, and they show good agreement with the experimental data. Finally, the amount of photoluminescence from each spin is assumed to be uniform. This assumption is not necessarily true considering the distribution of optical excitation power and different collection efficiency for each center.

In Fig. 10, the inhomogeneity of the barrel-shaped coil and planar system are shown as red circles and black squares, respectively. The width $\Delta\zeta$ obtained from the barrel-shaped coil is generally smaller than that of the planar system. This demonstrates the remarkable improvement of the magnetic field homogeneity for the barrel-shaped coil. At the same time, it becomes more prominent when measured away from the edge of the structure, which is located 1.5 mm from the center. This is relevant when it is required to locate the diamond at the edge of the structure because of space limitations given by the PL collection system, such as when a compound parabolic concentrator is used.[4]

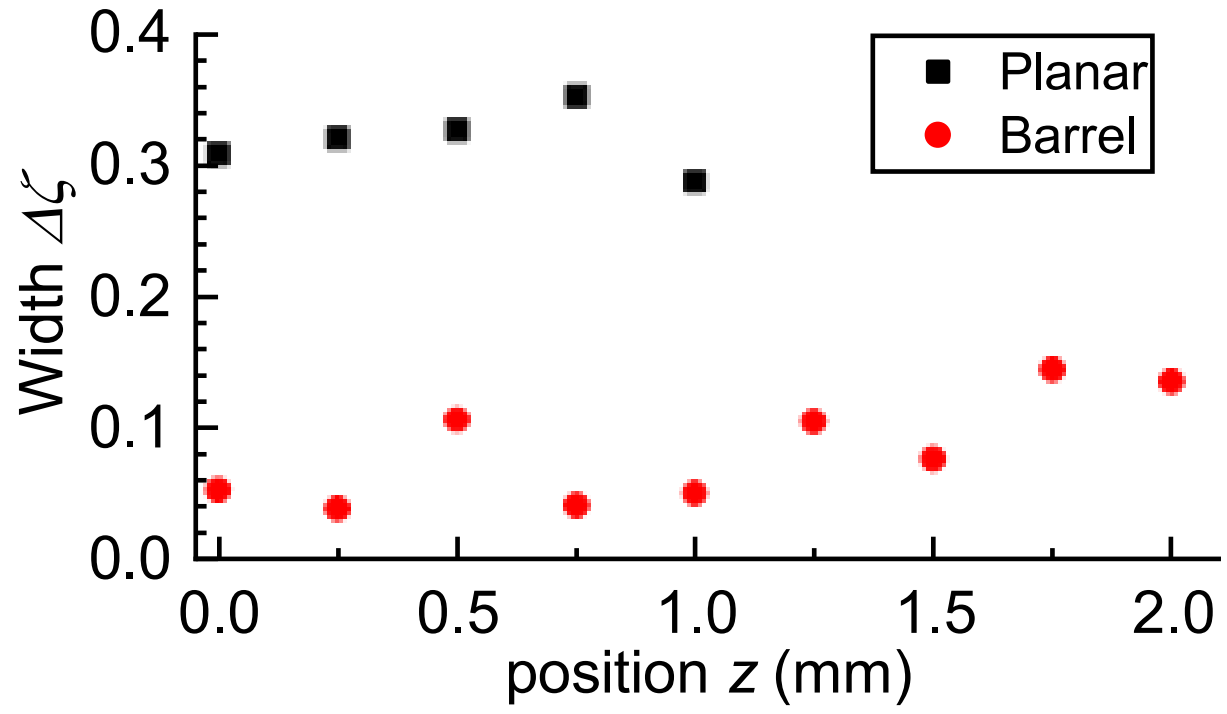

FIG. 10. Inhomogeneity of the planar system and barrel-shaped coil at different positions of the measurements. The results of Rabi oscillations are fit with exponentially decaying sinusoidal functions $S = Ae^{-t/T_\mathrm{R}} \cos \Omega_0 t$ and the inhomogeneity is evaluated by its width $\Delta\zeta = \frac{1}{\Omega_0 T_R}$. The width of the planar structure with different positions from the PCB board is plotted as black squares, while that of the barrel-shaped coil is plotted as red circles.

## V. CONCLUSIONS

By modeling the magnetic field distribution, in the so-called two-dimensional field-forming system (based on PCB technology),[11] a magnetic field inhomogeneity of less than 0.5% is achieved in a very small region, namely at |z|<0.075 mm (see Fig. 1 and Table I geometry A, Fig. 2). The use of dielectric resonators[25-28] allows for greater uniformity of the magnetic field. Thus, a magnetic field inhomogeneity of less than 0.5% is achieved in a significantly larger region at $|z|$<0.35 mm (see Fig. 1 and Table I geometry B, Fig. 3). Approximately the same magnetic field inhomogeneity occurs in the field-forming system made of three turns of conductive tape (see Fig. 1 and Table I geometry C, Fig. 4). The use of wire coils makes it possible to reduce the inhomogeneity of the magnetic field to fractions of a percent in the region where $|z|$<0.5 mm. In this case, the use of Helmholtz coils proves to be very effective (see Fig. 1 and Table I geometry D, Fig. 5). However, as noted,[32] the use of coils with a large number of turns appears to be more effective for increasing the level of magnetic induction which is very important for reducing the required source power, especially for mobile applications. Multi-turn field-forming systems also make it possible to ensure optimal distribution of current density in conductors along the $z$-axis of the field formation system, which leads to increased uniformity of the magnetic field. To investigate this, we carried out mathematical modeling of the proposed configuration of a six-turn barrel-shaped inductor system (see Fig. 6(a)). We observed a magnetic field inhomogeneity of less than 0.5% in a relatively large region where $|z|<0.5$ mm (see Fig. 6(c)), and less than 0.1% in the region where $|z|<0.25$ mm (see Fig. 1 and Table I geometry E). This is approximately 2 times less than in the case of Helmholtz coils (see Fig. 1 and Table I geometry D). When the number of turns of the inductor is doubled (see Fig. 7(c)), the magnetic field homogeneity remains approximately the same as that of 2- and 3-turn coils (see Fig. 1 and Table I geometry F, Fig. 7). It should be considered that the inductance of multi-turn inductors increases significantly with the number of turns which could reduce the inductor current, making matching not possible. However, this effect can be overcome by connecting groups of turns in parallel to the feeder, as is done in the proposed six- and twelve-turn field-forming systems.

Investigations of the non-uniformity of the magnetic field generated by coils with equal turn diameters are described elsewhere.[32] The same study compares the levels of non-uniformity of the magnetic field in a coil and in a Helmholtz coil system. As follows from,[32] Helmholtz coil provides a much more uniform magnetic field than the regular coils with equal turn diameters. In Soshenko, et al.,[45] experimentally obtained Rabi oscillations observed in the ensemble of NV centers placed in the field-forming system of Helmholtz coil are presented. As can be seen from a comparison of the attenuation degree of Rabi oscillations given in Soshenko, et al.[45] (see Fig. 2d) and those obtained in this work for a barrel-shaped inductor (see Fig. 9a), the attenuation degree of Rabi oscillations for the ensemble of NV centers in the first case is approximately 20% greater than in the second. Thus, it can be concluded that the proposed barrel-shaped inductor provides a greater degree of uniformity of the magnetic field than the Helmholtz coil. Therefore, it provides an even greater degree of homogeneity than coils with equal turn diameters.

After studying the magnetic field distributions obtained by mathematical modeling, the configuration and geometric parameters of a six-turn barrel-shaped field-forming system were proposed. The methods we developed to manufacture such a system have been presented and discussed (see Fig. 8). As shown in the experimental studies, the proposed field-forming system made it possible to significantly increase the degree of uniformity of the magnetic field compared to the existing field-forming systems (see Fig. 9).

## SUPPLEMENTARY MATERIAL

See the supplementary material for detailed information on the plots and fits of Rabi oscillations in chapter III of the main text.

## ACKNOWLEDGMENTS

We acknowledge the support of BMBF by future cluster QSENS (projects 03ZU1110GB QSCALE, 03ZU1110JA QSPACE) and projects DE-Brill 13N16207, 13N16215 SPINNING , 13N16463 DIAQNOS, 13N16707 quNV2.0, QRX and 13N15375 Quamopolis, DLR via project 50WM2170 QUASIMODO, Deutsche Forschungsgemeinschaft (DFG) via Projects No.386028944, No.445243414, and No.387073854 and Excellence Cluster POLiS, EU HORIZON program via projects QuMicro, SPINUS, C-QuENS, QCIRCLE and FLORIN, European Research Council (ERC) via Synergy grant HyperQ (Grant No. 856432) and Carl-Zeiss-Stiftung via the Center of Integrated Quantum Science and technology (IQST) and project Utrasens-Vir.

## AUTHOR DECLARATIONS

### Conflict of Interest

The authors have no conflicts to disclose.

### Author Contributions

**O. Rezinkin**: Conceptualization (lead); Resources(lead); Data curation (lead); Validation (equal); Writing – original draft (equal); Writing – review & editing (lead). **M. Rezinkina**: Conceptualization (equal); Software (lead); Data curation (equal); Formal analysis(equal); Validation (equal); Writing – original draft (lead); Writing – review & editing (equal). **T. Kitamura**: Data curation (equal); Investigation(lead); Formal analysis(equal); Validation (lead); Writing – original draft (equal); Writing – review & editing (equal). **R. Paul**: Data curation (supporting); Validation (supporting); Writing – review & editing (equal). **F. Jelezko**: Project administration (lead); Supervision (lead); Methodology (lead); Conceptualization (lead); Data curation (equal).

## DATA AVAILABILITY

The data supporting this study's findings are available from the corresponding authors upon reasonable request.

## APPENDIX A: RABI MEASUREMENT

Before conducting Rabi measurements, the continuous wave optically detected magnetic resonance measurements are performed to determine the resonant frequency at each configuration. See the supplementary information for their details. The input frequency is the central frequency when fit with a Gaussian function. 2.838 GHz is used for the barrel system, while 2.995 GHz is used for the planar system.

Rabi oscillations of NV centers are measured by collecting the spin-dependent red-shifted photoluminescence (PL). The pulse sequences are shown in Fig. 11. Firstly, the spin states are initialized by applying a green laser to the NV centers for $t_L = 2$ ms. Then, the resonant MW fields are irradiated for a duration $\tau$. Finally, the spin states are read out by averaging the amount of photoluminescence. The bare Rabi signals are accompanied by the baseline shifts, which we attribute to the fluctuation of the laser power. To remove this noise, similar measurements without MW irradiation are conducted. The signals shown in the main text are obtained by subtracting these reference signals from the bare Rabi signals.

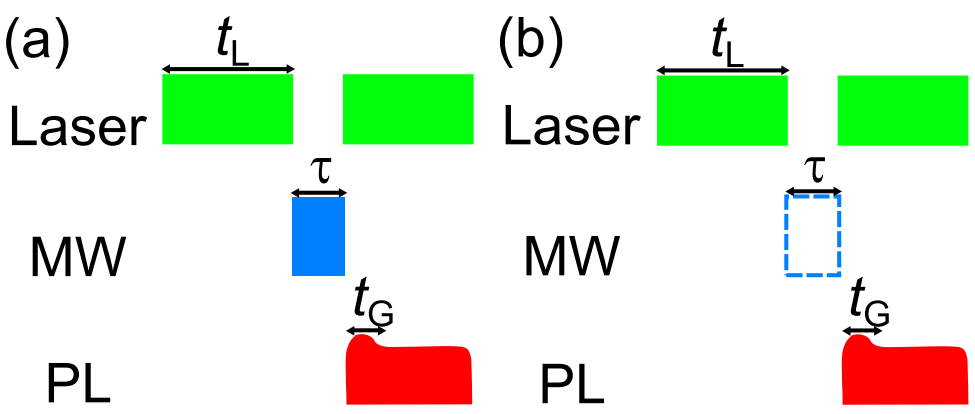

FIG. 11. (a) – pulse sequence for Rabi measurements with NV centers. After initializing the NV spin states with a green laser for time $t_L$, the resonant MW is applied for the duration $\tau$. Their spin states are read out by averaging the initial spin-dependent photoluminescence (PL) gated for duration $t_G$. (b) – pulse sequence for referencing. The bare Rabi signals have baseline shifts, which we attribute to the fluctuation of the laser power. The signal of this reference measurement is used to subtract them.

**APPENDIX B: Experimental setup**

The diamond sample (Element 6, DNV-B1) is cut down by a laser cutter (Syntek). The density of NV is approximately 300 ppb from the specification. According to the microscope measurements, its measured length is 940 μm.

The schematic of the optical setup is shown in Fig. 12. Continuous wave green laser is output from a diode pump solid-state laser (Spectra Physics, Millennia eV5) with 500 mW power at a wavelength of 532 nm, and its output is tunable with a half-wave plate (Thorlabs, WPH10M-532) and a polarizing beam splitter (Thorlabs, PBS251), where the rest is blocked by a beam blocker (Thorlabs, BT620). It is focused by a plano-convex lens (Thorlabs, AC-254-300-A-ML) onto an acoustic-optic modulator (AOM) (Crystal Technology 3250-220) for pulsing after its polarization is adjusted by another half-wave plate (Thorlabs, WPH10M-532). The beam is collimated by another plano-convex lens (Thorlabs, AC-254-300-A-ML), and its diameter is adjusted by a beam expander (Thorlabs, ZBE22). Then, it is focused on the diamond with a diameter of 45 μm. The PL is collected using a Gradient-Index (GRIN) lens (Thorlabs, G2P10) attached to a compound parabolic concentrator (CPC) (Edmund, #17-709) with optical glue (M-GLASS). After passing through a long-pass filter (AHF Analysentechnik, BLP01-633R-25), it is detected with a photodetector (Thorlabs, PDA100A2), followed by analog-digital conversion with a digitizer (Spectrum Instrumentation, M4i.4420-x8). The beam diameter is checked by a camera behind the dichroic mirror (Thorlabs, DMLP650). The photoluminescence is collected by the lens and imaged by another lens onto a camera (Thorlabs, CS165MU).

Two three-axis stages adjust the diamond's relative position to the PCB board. The NV diamond is attached to the top of the GRIN lens, which can be adjusted by a three-axis stage connected at the CPC. The PCB is fixed to another stage, which is adjusted along the axial direction by a micrometer in the experiments (Fig.12).

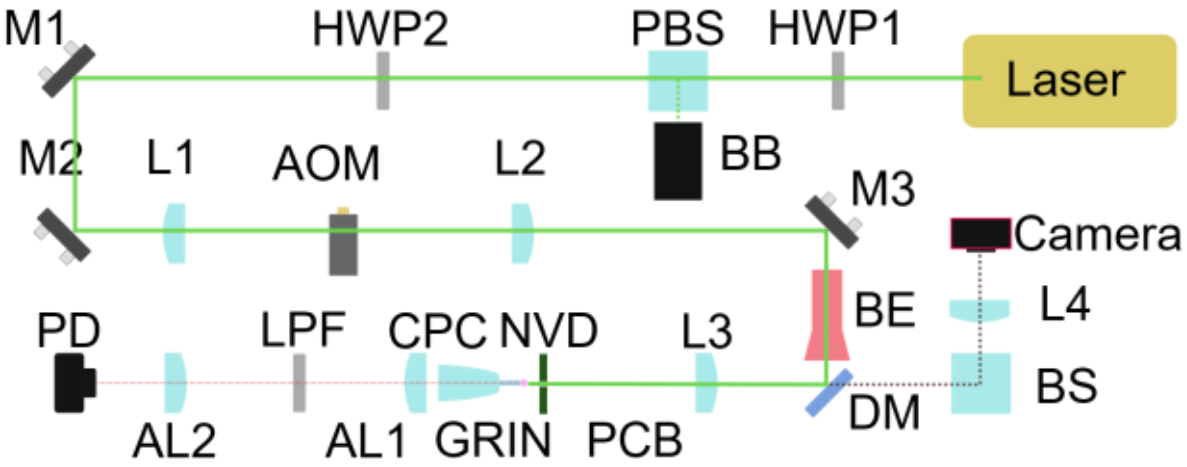

FIG. 12. Schematic of the optical setup. Laser power is finely tunable with a half-wave plate (HWP1). A polarizing beam splitter (PBS) and a beam blocker (BB). After adjusting the polarization with another half-wave plate (HWP), the laser is focused on an acoustic-optic modulator (AOM) with a plano-convex lens (L1). The beam is collimated with another lens (L2) and passes through a beam expander (BE). This beam is reflected at a dichroic mirror (DM) and focused on the NV diamond (NVD) passing through the hole of a printed circuit board (PCB). The red-shifted photoluminescence is collected through a gradient-index (GRIN) lens glued to a compound parabolic concentrator (CPC). It is led to the photodetector (PD) with aspheric lenses (AL1, AL2) and a long-pass filter (LPF). The focus of the laser is checked by a camera behind the dichroic

mirror. Light is collected by the lens (L3) and transmitted behind the dichroic mirror. It is imaged through another lens (L4) onto a camera after being reflected at a beam splitter (BS).

A neodymium magnet is placed around the diamond sample to apply the static bias magnetic field. This is used to distinguish the four groups of NV centers with different Zeeman axes originating from the different relative directions of nitrogen and vacancy in the diamond lattice. MW signal with a peak-to-peak voltage of 200 mV is generated by an arbitrary waveform generator (Keysight, M8915A), which is amplified with a gain of 45 dB by a MW amplifier (Mini-Circuits, ZHL-16W-43-S+), and in turn, is fed into the SMA connector on the PCB.

## REFERENCES

[1]L. Rondin, J.-P. Tetienne, T. Hingant, J.-F. Roch, P. Maletinsky, and V. Jacques, "Magnetometry with nitrogen-vacancy defects in diamond," Reports on Progress in Physics **77**, 056503 (2014).

[2]F. Dolde, H. Fedder, M. W. Doherty, T. Nöbauer, F. Rempp, G. Balasubramanian, T. Wolf, F. Reinhard, L. C. Hollenberg, F. Jelezko, and J. Wrachtrup, "Electric-field sensing using single diamond spins," Nature Physics **7**, 459–463 (2011).

[3]S. Choe, J. Yoon, M. Lee, J. Oh, D. Lee, H. Kang, C.-H. Lee, and D. Lee, "Precise temperature sensing with nanoscale thermal sensors based on diamond NV centers," Current Applied Physics **18**, 1066–1070 (2018).

[4]T. Wolf, P. Neumann, K. Nakamura, H. Sumiya, T. Ohshima, J. Isoya, and J. Wrachtrup, "Subpicotesla Diamond Magnetometry," Phys. Rev. X **5**, 041001 (2015)

[5]X. Zhu, S. Saito, A. Kemp, K. Kakuzanagi, S. Karimoto, H. Nakano, W. J. Munro, Y. Tokura, M. S. Everitt, K. Nemoto, M. Kasu, N. Mizuochi, and K.Semba, "Coherent coupling of a superconducting flux qubit to an electron spin ensemble in diamond," Nature **478**, 221–224 (2011).

[6]F. Jelezko, and J. Wrachtrup, "Single defect centres in diamond: A review," Physica Status Solidi (a) **203** (2006).

[7]C. Zhang, H. Yuan, N. Zhang, L. Xu, J. Zhang, B. Li, and J. Fang, "Vector Magnetometer Based on Synchronous Manipulation of Nitrogen-Vacancy Centers in All Crystal Directions," J. Phys. D. Appl. Phys. **51** (2018).

[8]K. Oshimi, Y. Nishimura, T. Matsubara, M. Tanaka, E. Shikoh, L. Zhao, Y. Zou, N. Komatsu, Y. Ikado, Y. Takezawa, E. Kage-Nakadai, Y. Izutsu, K. Yoshizato, S. Morita, M. Tokunaga, H. Yukawa, Y. Baba, Y. Teki, and M. Fujiwara, "Glass-Patternable Notch-Shaped Microwave Architecture for on-Chip Spin Detection in Biological Samples," Lab Chip **22**, 2519 (2022).

[9]M. Chipaux, A. Tallaire, J. Achard, S. Pezzagna, J. Meijer, V. Jacques, J. F. Roch, and T. Debuisschert, "Magnetic Imaging with an Ensemble of Nitrogen Vacancy-Centers in Diamond," Eur. Phys. J. D **69**, 1 (2015).

[10]A. Horsley, P. Appel, J. Wolters, J. Achard, A. Tallaire, P. Maletinsky, and P. Treutlein, "Microwave Device Characterization Using a Widefield Diamond Microscope," Phys. Rev. Appl. **10** (2018).

[11]O. R. Opaluch, N. Oshnik, R. Nelz, and E. Neu, "Optimized Planar Microwave Antenna for Nitrogen Vacancy Center Based Sensing Applications," Nanomaterials **11** (2021).

[12]A. Savitsky, J. Zhang, and D. Suter, Variable Bandwidth, "High Efficiency Microwave Resonator for Control of Spin-Qubits in Nitrogen-Vacancy Centers," Rev. Sci. Instrum. **94** (2023).

[13]Y. Guo, J. Zhao, C. Weng, S. Lin, Y. Yang, W. Zhu, L. Lou, and G. Wang, "Robust Diamond-Embedded Microwave Antenna for Optimizing Quantum Sensing Using Nitrogen-Vacancy Center Ensembles," Appl. Phys. Lett. **123** (2023).

[14]K. Bayat, J. Choy, A. V. Shneidman, S. Meesala, M. F. Baroughi, and M. Loncar, "Uniform and Large Volume Microwave Magnetic Coupling to NV Centers in Diamond Using Split Ring Resonators," Opt. InfoBase Conf. Pap. (2014).

[15]Y. Gao, H. Guo, H. Wen, Z. Li, Z. Ma, J. Tang, and J. Liu, "CSRR Structure Design for NV Spin Manipulation with Microwave Strength and Fluorescence Collection Synchronous Enhancement," Materials (Basel). **16** (2023).

[16]Y. Yang, Q. Wu, Y. Wang, W. Chen, Z. Yu, X. Yang, J.-W. Fan, and B. Chen, "Tunable Double Split-Ring Resonator for Quantum Sensing Using Nitrogen-Vacancy Centers in Diamond," Opt. Contin. **2**, 1426 (2023).

[17]N. Zhang, C. Zhang, L. Xu, M. Ding, W. Quan, Z. Tang, and H. Yuan, "Microwave Magnetic Field Coupling with Nitrogen-Vacancy Center Ensembles in Diamond with High Homogeneity," Appl. Magn. Reson. **47**, 589 (2016).

[18]K. Sasaki, Y. Monnai, S. Saijo, R. Fujita, H. Watanabe, J. Ishi-Hayase, K. M. Itoh, and E. Abe, Broadband, "Broadband, large-area microwave antenna for optically detected magnetic resonance of nitrogen-vacancy centers in diamond," Rev. Sci. Instrum. **87**, 1 (2016).

[19]L. Qin, Y. Fu, S. Zhang, J. Zhao, J. Gao, H. Yuan, Z. Ma, Y. Shi, and J. Liu, "Near-Field Microwave Radiation Function on Spin Assembly of Nitrogen Vacancy Centers in Diamond with Copper Wire and Ring Microstrip Antennas," Jpn. J. Appl. Phys. **57** (2018).

[20]Z. Ma, D. Zheng, Y. Fu, H. Yuan, W. Guo, J. Zhao, B. Li, Y. Shi, X. Zhang, and J. Liu, "Efficient Microwave Radiation Using Broadened-Bandwidth Coplanar Waveguide Resonator on Assembly of Nitrogen-Vacancy Centers in Diamond," Jpn. J. Appl. Phys. **58** (2019).

[21]M. Zhao, Q. Lin, L. Zhu, L. Zhao, and Z. Jiang, "Antenna for Microwave Manipulation of NV Colour Centres," Micro Nano Lett. **15**, 793 (2020).

[22]Z. Li, Z. Li, Z. Shi, H. Zhang, Y. Liang, and J. Tang, "Design of a High-Bandwidth Uniform Radiation Antenna for Wide-Field Imaging with Ensemble NV Color Centers in Diamond," Micromachines **13** (2022).

[23]S. Mahtab, P. Milas, D. T. Veal, M. G. Spencer, and B. Ozturk, "High Efficiency Radio Frequency Antennas for Amplifier Free Quantum Sensing Applications," Rev. Sci. Instrum. **94** (2023).

[24]Y. Chen, T. Li, D. Wang, B. Lu, G. Chai, and J. Tian, "Compact Multipass-Laser-Beam Antenna for NV Sensor Sensitivity Enhancement," Opt. Express **31**, 33123 (2023).

[25]J. F. Barry, M. H. Steinecker, S. T. Alsid, J. Majumder, L. M. Pham, et al., "Sensitive AC and DC Magnetometry with Nitrogen-Vacancy Center Ensembles in Diamond," e-Print: 2305.06269 (2023).

[26]E. R. Eisenach, J. F. Barry, L. M. Pham, R. G. Rojas, D. R. Englund, and D. A. Braje, "Broadband Loop Gap Resonator for Nitrogen Vacancy Centers in Diamond," Rev. Sci. Instrum. **89**, 1 (2018).

[27]V. Yaroshenko, V. Soshenko, V. Vorobyov, S. Bolshedvorskii, E. Nenasheva, I. Kotel'nikov, A. Akimov, and P. Kapitanova, "Circularly Polarized Microwave Antenna for Nitrogen Vacancy Centers in Diamond", Rev. Sci. Instrum. **91** (2020).

[28]H. H. Vallabhapurapu, J. P. Slack-Smith, V. K. Sewani, C. Adambukulam, A. Morello, J. J. Pla, and A. Laucht, "Fast Coherent Control of a Nitrogen-Vacancy-Center Spin Ensemble Using a Dielectric Resonator at Cryogenic Temperatures," Phys. Rev. Appl. **16**, 1 (2021).

[29]Y. Chen, H. Guo, W. Li, D. Wu, Q. Zhu, B. Zhao, L. Wang, Y. Zhang, R. Zhao, W. Liu, F. Du, J. Tang, and J. Liu, "Large-Area, Tridimensional Uniform Microwave Antenna for Quantum Sensing Based on Nitrogen-Vacancy Centers in Diamond," Appl. Phys. Express **11** (2018).

[30]Y. Chen, T. Li, G. Chai, D. Wang, B. Lu, A. Guo, and J. Tian, "Enhancing Spin-Based Sensor Sensitivity by Avoiding Microwave Field Inhomogeneity of NV Defect Ensemble," Nanomaterials **12** (2022).

[31]Y. Ben-Shalom, A. Hen, and N. Bar-Gill, "Modified Split Ring Resonators for Efficient and Homogeneous Microwave Control of Large

Volume Spin Ensembles," arXiv:2309.11130 (2023) IEEE Sensors J. 24, 20420 (2024).
[32]C. Park and D. Lee, "Design and Simulation of a Strong and Uniform Microwave Antenna for a Large Volume of Nitrogen-Vacancy Ensembles in Diamond," J. Korean Phys. Soc. **78**, 280 (2021).
[33]D. B. Bucher, D. P. L. Aude Craik, M. P. Backlund, M. J. Turner, O. Ben Dor, D. R. Glenn, and R. L. Walsworth, "Quantum Diamond Spectrometer for Nanoscale NMR and ESR Spectroscopy," Nat. Protoc. **14**, 2707 (2019).
[34]Y. Takemura, K. Hayashi, Y. Yoshii, M. Saito, S. Onoda, H. Abe, T. Ohshima, T. Taniguchi, M. Fujiwara, H. Morishita, I. Ohki, and N. Mizuochi, "Broadband Microwave Antenna for Uniform Manipulation of Millimeter-Scale Volumes of Diamond Quantum Sensors," J. Appl. Phys. **132** (2022).
[35]J. M. Schloss, J. F. Barry, M. J. Turner, and R. L. Walsworth, "Simultaneous Broadband Vector Magnetometry Using Solid-State Spins," Phys. Rev. Applied **10**, 034044 (2018).
[36] K. Yahata, Y. Matsuzaki, S. Saito, H. Watanabe, and J. Ishi-Hayase, "Demonstration of vector magnetic field sensing by simultaneous control of nitrogen-vacancy centers in diamond using multi-frequency microwave pulses," Appl. Phys. Lett. **114**, 022404 (2019).
[37]J-M. Cai, B. Naydenov, R. Pfeiffer, L. P. McGuinness, K. d. Jahnke, F. Jelezko, M. B. Plenio, and A. Retzker, "Robust dynamical decoupling with concatenated continuous driving," New J. Phys. **14**, 113023 (2012).
[38]A. Stark, M. Aharon, T. Unden, D. Louzon, A. Huck, A. Retzker, U. L. Andersen, and F. Jelezko, "Narrow-bandwidth sensing of high-frequency fields with continuous dynamical decoupling," Nat. Commun. **8**, 1105 (2017).
[39]I. Niemeyer, J. H. Shim, J. Zhang, D. Suter, T. Taniguchi, T. Teraji, H. Abe, S. Onoda, T. Yamamoto, and T. Ohshima, "Broadband excitation by chirped pulses: application to single electron spins in diamond," New. J. Phys. **15**, 033027 (2013).
[40]F. M. Stürner, A. Brenneis, J. Kassel, U. Wostradowski, R. Rölver, T. Fuchs, K. Nakamura, H. Sumiza, S. Onoda, J. Isoza, and F. Jelezko, "Compact integrated magnetometer based on nitrogen-vacancy centres in diamond," Diam. Relat. Mater. **93**, 59 (2019).
[41]F. M. Stürner, A. Brenneis, T. Buck, J. Kassel, R. Rölver, U. Wostradowski, T. Fuchs, A. Savitskz, D. Suter, J. Grimmel, S. Hengesbach, Michael Förtsch, K. Nakamura, H. Sumiza, S. Onoda, J. Isoza, and F. Jelezko, "Integrated and Portable Magnetometer Based on Nitrogen-Vacancy Ensembles in Diamond," Adv. Quantum Technol. **4**, 20000111 (2021).
[42]C. Bowick, RF circuit design, Boston: Elsevier, 2008.
[43]H. De Raedt, B. Barbara, S. Miyashita, K. Michielsen, S. Bertaina, and S. Gambarelli, "Quantum Simulations and Experiments on Rabi Oscillations of Spin Qubits: Intrinsic vs Extrinsic Damping," Phys Rev B Condens Matter Mater Phys **85** (2012).
[44]F. Jelezko, T. Gaebel, I. Popa, A. Gruber, and J. Wrachtrup, "Observation of Coherent Oscillations in a Single Electron Spin," Phys. Rev. Lett. **92**, 076401 (2004).
[45]V. V. Soshenko, O. R. Rubinas, V. V. Vorobyov, S. V. Bolshedvorskiia, P. V. Kapitanovac, V. N. Sorokina, and A. V. Akimova, "Microwave Antenna for Exciting Optically Detected Magnetic Resonance in Diamond NV Centers," Bull. Lebedev Phys. Inst. **45**, 237 (2018).

## Supplementary Material for
# Uniform microwave field formation for control of ensembles of negatively charged nitrogen vacancy in diamond

Oleg Rezinkin,[1,2] Marina Rezinkina,[1,2*] Takuya Kitamura,[1] Rajan Paul[1] and Fedor Jelezko[1]

1 Ulm University, Ulm, 89081, Germany
2 National Technical University "Kharkiv Polytechnic Institute", Kharkiv, 61002, Ukraine

This supplementary material includes the details of Rabi measurements in chapter III of the main text. The first section describes the continuous wave optically detected magnetic resonance(CW ODMR) measurements performed to find the resonant frequencies of the NV centers under bias magnetic field. The second section shows the noise level of microwave signal to exclude that the decay in Rabi oscillations does not result from the fluctuation of the amplitude. The third section summarizes the Rabi measurements carried out in the main text.

## 1. ODMR

Figure S1 shows the ODMR spectrum obtained (a) when the barrel structure was tested and (b) when the planar system was tested. Due to the four groups of Zeeman axes resulting from the direction of nitrogen and vacancy in the diamond lattice, four splittings of the dips from the zero field splitting $D = 2.87$ GHz could be observed in the spectrum. In the measurements with the barrel structure, the transition around 2.84 GHz is used. Fitting with a Gaussian function yields a central frequency of 2.838 GHz and a width of 4.732 MHz, which is represented by red curve in the figure. By attributing the Zeeman shift to the magnetic field parallel to the Zeeman axis, the resonant frequency reads $\omega = D - \gamma B_z$. Approximately 1.5 mT is applied to the NV center. Likewise, the transition around 2.95 GHz is used in the measurements with the planar system. Fitting with a Gaussian function yields a central frequency of 2.995 GHz and a width of 7.401 MHz. By use of $\omega = D + \gamma B_z$, the applied bias filed is approximated as 4.5 mT. The validity of the chosen frequencies is discussed later in Section 3.

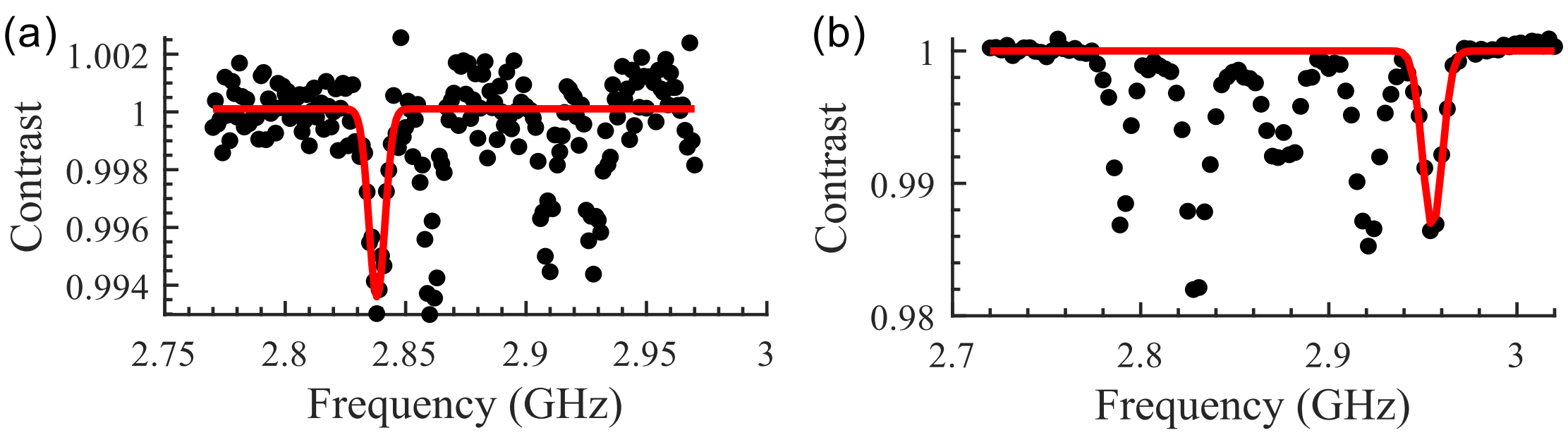

FIG. S1. CW ODMR measurements and their fits to the Gaussian function: (a) with the barrel inductor; (b) with the planer system.

## 2. Noise level of microwave signal

The noise level of the microwave signal input to the PCB board is measured with an oscilloscope (Keysight, MSOS404A). A sinusoidal waveform is generated in the AWG and amplified with the amplifier as described in the Appendix B in the main article. This is connected to an attenuator (Mini-circuits, BW-N40W50+) and fed into the oscilloscope. The amplitude error is estimated as the ratio of the standard deviation to the mean value of the peak-to-peak voltage. Likewise, the frequency error is also estimated. Table S1 summarizes these values. The amplitude error is found to be less than 1 %, which suggests that the fluctuation of amplitudes in the temporal averaging is not the main cause of the decays observed in the Rabi oscillations.

TABLE S1. List of errors in the microwave signal

| Frequency (GHz) | Frequency error (%) | Amplitude error (%) |
|---|---|---|
| 2.838 | 0.41 | 0.67 |
| 2.995 | 0.41 | 0.68 |

**3. Summary of the Rabi measurements**

The Rabi oscillations measured with the barrel system and the planar system are summarized in fig. S2 and fig. S3, respectively. They are fit with the exponentially decaying sinusoidal function: $f(t) = A\exp\left(-\frac{t}{T}\right)\cos(2\pi f t + q) + B$. Table S2 and S3 summarize the fitting parameters for the barrel system and the planar system, respectively.

The decay of Rabi oscillations can be caused by several mechanisms. In the main article, the homogeneity of the magnetic field created by the system is estimated by attributing the decay solely to it. We consider this approach valid for the following reasons. Let us consider a two level system with energy level of $\omega_0$ driven by control field with frequency of $\omega$ and with Rabi frequency of $\Omega$. We model this with the Hamiltonian $H = \frac{\omega_0}{2}\sigma_z + \Omega\cos\omega t\,\sigma_x$, for simplicity. In the rotating frame with rotating wave approximation, this reads $H' = \frac{\delta\omega}{2}\sigma_z + \frac{\Omega}{2}\sigma_x$, where $\delta\omega = \omega_0 - \omega$ is the detuning. In each repetition, signal represents the spatial ensemble of the evolutions of systems weighted by the amount of their photoluminescence. Averaging by repeated measurements gives the temporal ensemble of the signal. Under this argument, the shape of the signal is attributed to the distribution of the detunings and the driving strength among this ensemble. Fluctuation of the microwave amplitude can trigger the decay of Rabi oscillation,[1] but the low amplitude noise level suggests that this is not the main cause of the decay. Inhomogeneous broadenings can also cause the decay when it is larger than the Rabi frequency. This effect could occur more likely with the barrel system because the Rabi frequency is less than 1 MHz, which is smaller than that with the planar system (more than 1 MHz). The good fitting to the sinusoidal single exponential function suggests the less effect of the inhomogeneous broadening because the Rabi oscillation with the inhomogeneous broadening much larger than the Rabi frequency should behave differently.[2,3] Thus, we attribute the shape of Rabi oscillation mainly to the inhomogeneity of the driving field. We consider that the anomalous decay of the Rabi oscillation with the planar system results from the distribution of the magnetic field originating from the planar structure.

At two different frequencies, there might be changes in the amplitude and the direction of the bias magnetic field and the MW magnetic field. In this study, they are assumed to be homogeneous except for the amplitude of the MW magnetic field. The above argument indicates that for simple estimation of large inhomogeneities, the effect of inhomogeneity is dominant in the decay of Rabi oscillations when the Rabi frequency is relatively higher than the inhomogeneous broadening.

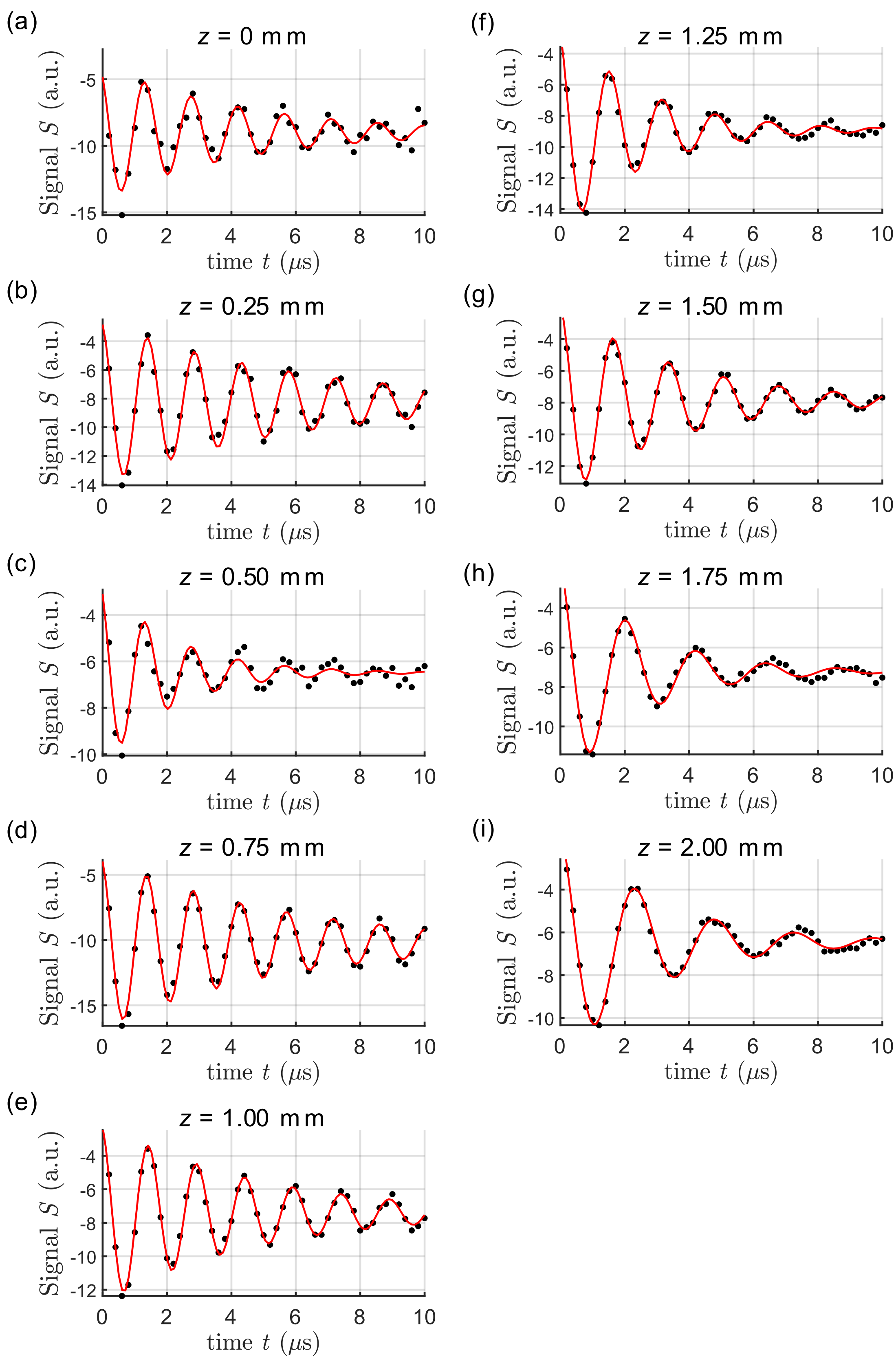

FIG. S2. Rabi oscillations with the barrel inductor at different $z$ positions: (a) $z$ = 0 mm; (b) $z$ = 0.25 mm; (c) $z$ = 0.50 mm; (d) $z$ = 0.75 mm; (e) $z$ = 1.00 mm; (f) $z$ = 1.25 mm; (g) $z$ = 1.50 mm; (h) $z$ = 1.75 mm; (i) $z$ = 2.00 mm.

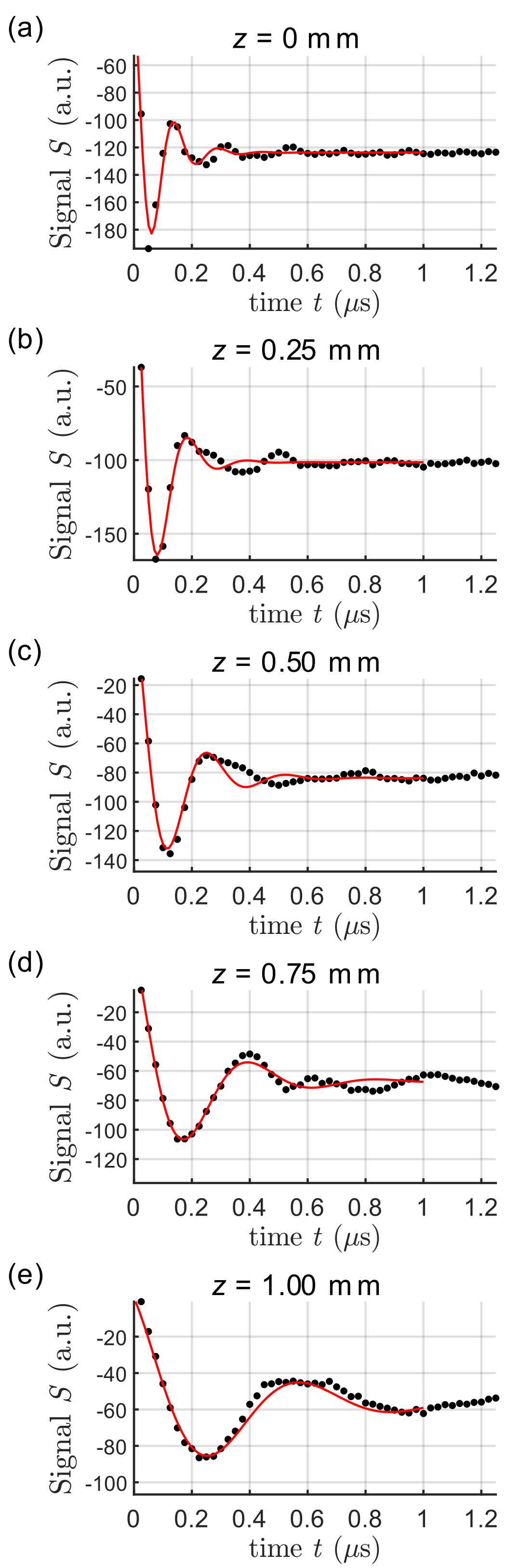

FIG. S3. Rabi oscillations with the planar system at different $z$ position: (a) $z = 0$ mm; (b) $z = 0.25$ mm; (c) $z = 0.50$ mm; (d) $z = 0.75$ mm; (e) $z = 1.00$ mm.

TABLE S2. List of fitting parameters for the barrel inductor

| $z$(mm) | $A$ | $B$ | $T$(μs) | $f$(kHz) | $q$ |
|---|---|---|---|---|---|
| 0 | 5.07 | -8.98 | 4.37 | 690 | 0.61 |
| 0.25 | 5.69 | -8.26 | 6.04 | 683 | 0.30 |
| 0.50 | 4.00 | -6.48 | 2.17 | 689 | 0.56 |
| 0.75 | 6.60 | -10.21 | 5.61 | 690 | 0.33 |
| 1.00 | 5.43 | -7.42 | 4.74 | 667 | 0.33 |
| 1.25 | 6.92 | -8.90 | 2.50 | 607 | 0.43 |
| 1.50 | 6.24 | -7.87 | 3.56 | 583 | 0.22 |
| 1.75 | 6.08 | -7.21 | 2.39 | 458 | 0.36 |
| 2.00 | 5.61 | -6.46 | 2.89 | 404 | 0.31 |
| 2.25 | 3.53 | -5.08 | 2.58 | 283 | 0.46 |
| 2.50 | 2.66 | -3.97 | 3.14 | 222 | 0.29 |

TABLE S3. List of fitting parameters for the planar system

| $z$(mm) | $A$ | $B$ | $T$(ns) | $f$(MHz) | $q$ |
|---|---|---|---|---|---|
| 0 | 133.5 | -123.7 | 78.6 | 6.54 | 0.36 |
| 0.25 | 191.4 | -101.5 | 77.9 | 4.83 | 0.31 |
| 0.50 | 121.4 | -83.7 | 132.2 | 3.68 | 0.17 |
| 0.75 | 98.9 | -67.1 | 199.5 | 2.26 | 0.37 |
| 1.00 | 61.8 | -56.7 | 349.3 | 1.58 | 0.35 |

**REFERENCES**

[1]F. Jelezko *et al*., Phys. Rev. Lett. **92**, 7 (2004).
[2]E. Baibekov *et al*., J. Mag. Res. **209**, 61 (2011).
[3]H. De Raedt *et al*., Phys. Rev. Lett. **85**, 014408 (2012).